\documentclass[12pt]{iopart}

\usepackage{graphicx}
\usepackage{comment}
\usepackage[T1]{fontenc}
\usepackage{dsfont}
\usepackage{xcolor}

\graphicspath{{../},{figures/}}

\newcommand{\openone}{\mathds{1}}
\newcommand{\PT}{${\mathcal{PT}}$}

\newcommand{\LL}{\mathcal{L}}
\newcommand{\DD}{\mathcal{D}}

\newcommand{\rhot}{\hat{\rho}(t)}
\newcommand{\ket}[1]{| #1 \rangle}
\newcommand{\bra}[1]{\langle #1 |}
\def \etal{\textit{et al.}}
\def \Vec#1{{\cal F}(#1)}
\def \Invec#1{{\cal F}^{-1}(#1)}
\def\In{{\rm in}}
\def\Out{{\rm out}}
\def\Tr{{\rm tr}}

\newcommand{\eig}[1]{\hat{\rho}_{#1}}

\def\diag{{\rm diag}}

\begin{document}

\title{Experimental Liouvillian exceptional points in a quantum system
without Hamiltonian singularities}

\author{Shilan Abo$^{1}$, Patrycja Tulewicz$^{1}$, Karol Bartkiewicz$^{1}$,
\c{S}ahin K. \"Ozdemir$^{2}$, and Adam
Miranowicz$^{1}$\footnote{Author to whom any correspondence should
be addressed.}}

\address{$^{1}$ Institute of Spintronics and
Quantum Information, Faculty of Physics, Adam Mickiewicz
University, 61-614 Pozna\'{n}, Poland}

\address{$^{2}$ Department of Electrical and Computer Engineering, Saint Louis
University, St. Louis MO 63103, USA}

\ead{adam.miranowicz@amu.edu.pl}

\date{\today}

\begin{abstract}
Hamiltonian exceptional points (HEPs) are spectral degeneracies of
non-Hermitian Hamiltonians describing classical and semiclassical
open systems with losses and/or gain. However, this definition
overlooks the occurrence of quantum jumps in the evolution of open
quantum systems. These quantum effects are properly accounted for
by considering quantum Liouvillians and their exceptional points
(LEPs). Specifically, an LEP corresponds to the coalescence of two
or more eigenvalues and the corresponding eigenmatrices of a given
Liouvillian at critical values of external parameters [Minganti
\emph{et al.}, Phys. Rev. A {\bf 100}, 062131 (2019)]. Here, we
explicitly describe how standard quantum process tomography, which
reveals the dynamics of a quantum system, can be readily applied
to detect and characterize quantum LEPs of quantum non-Hermitian
systems. We conducted experiments on an IBM quantum processor to
implement a prototype model with one-, two-, and three qubits
simulating the decay of a single qubit through competing channels,
resulting in LEPs but not HEPs. Subsequently, we performed
tomographic reconstruction of the corresponding experimental
Liouvillian and its LEPs using both single- and two-qubit
operations. This example underscores the efficacy of process
tomography in tuning and observing LEPs even in the absence of
HEPs.
\end{abstract}

\maketitle

\section{Introduction}

Systems with dissipation and/or amplification can be described by
non-Hermitian Hamiltonians (NHHs) whose eigenvalues are either
real or complex conjugate pairs depending on whether the system is
operated in the exact or broken parity-time (\PT) symmetric phase,
respectively~\cite{Bender1998,Bender2003,Bender2007}. Over the
past two decades, \PT-symmetric systems have evolved from a
mathematical curiosity to a powerful resource for controlling
electromagnetic waves and their interactions with matter by
judiciously engineering loss-imbalance in passive (i.e., without
amplification) non-Hermitian systems, and dissipation vs
amplification rates in active non-Hermitian systems, as well as
dissipation vs the coupling strength between
subsystems~\cite{MoiseyevBook, El-Ganainy2018}. Early
demonstrations involved optical~\cite{Guo2009, Ruter2010,
Regensburger2012}, electronic~\cite{Schindler2011},
plasmonic~\cite{Benisty2011}, metamaterial~\cite{Kang2013,
Fleury2014, Sun2014}, optomechanical, and acoustic~\cite{Zhu2014,
Jing2014, Fleury2015} systems, before further expanding to include
other platforms~\cite{Gao2015, Peng2016, Ding2021, Naghiloo2019,
Chen2021, Chen2022}).

The exact and broken \PT-symmetric phases are separated by the
so-called Hamiltonian exceptional points (HEPs), where two or more
of the eigenvalues of the effective NHH describing a given system,
and their associated eigenvectors, become degenerate, leading to
dimensionality reduction~\cite{Ozdemir2019, Miri2019, Parto2021}.

A plethora of intriguing properties of such systems induced or
enhanced at HEPs (or near them) have been predicted, including
stimulated~\cite{Feng2014, Hodaei2014, Peng2014a,
Brandstetter2014, Arkhipov2019, Arkhipov2020a} and
spontaneous~\cite{Pick2017} emission, chirality~\cite{Heiss2001,
Dembowski2003, Sweeney2019}, unidirectional
invisibility~\cite{Lin2011}, control of whispering-gallery
microcavities~\cite{Chang2014, Peng2014b}, an exceptional Kerr
effect~\cite{Perina2019} and related exceptional photon
blockade~\cite{Huang2019}, or the generation of higher-order
HEPs~\cite{Hodaei2017, Jing2017, Arkhipov2020b, Arkhipov2021}. The
existence of HEPs in the absence of the \PT-symmetry was studied
in Ref.~\cite{Lange2020}.

While effective NHHs and HEPs are sufficient to describe coherent
nonunitary evolution of the dynamics of classical and
semiclassical systems, they fell short in describing the evolution
of quantum systems involving quantum jumps and associated noise.
To address this shortcoming, quantum Liouvillian exceptional
points (LEPs) were introduced as degeneracies of quantum
Liouvillian superoperators associated with their coalescing
eigenvalues and eigenvectors~\cite{Minganti2019}. LEPs are a
natural generalization of HEPs by including quantum jumps to
provide a consistent description of decoherence and noise in open
quantum systems compatible with the canonical commutation
relations. Indeed, LEPs depend not only on a continuous nonunitary
dissipation of a given system (as described by NHHs), but also on
quantum jumps; this is contrary to HEPs which are not affected by
them, so in that sense can be considered semiclassical or even
classical. The connection between HEPs and LEPs can be
demonstrated by postselecting quantum trajectories following the
hybrid-Liouvillian formalism~\cite{Minganti2020}. Recent
experiments with a single three-level transmon~\cite{Chen2021,
Chen2022} and a single three-level trapped ion~\cite{Zhang2022,
Bu2023} have indicated the importance of LEPs by revealing the
pivotal significance of quantum jumps in generalizing the
applications of classical non-Hermitian systems to open quantum
systems. These applications encompass advanced techniques such as
precise sensing and control of quantum circuits~\cite{Chen2021},
dynamical manipulation of quantum thermal
machines~\cite{Khandelwal2021}, and specifically the operation of
quantum heat engines~\cite{Bu2023}, all exploiting the unique
properties of LEPs. The formalism of LEPs is based on the Lindblad
master equation, so relies on the standard quantum mechanics,
where there is no need for calculating a system-dependent
metric~\cite{Ju2019, Ju2022}, thus preventing the apparent
violation of the no-go theorems.

Although the complex spectra of Liouvillians have been analyzed
previously (see, e.g., Refs.~\cite{Rivas2011, Baumgartner2008a,
Baumgartner2008b, Albert2014, Minganti2018} and references
therein), interest in Liouvillian singularities, now termed LEPs
and Liouvillian diabolical points (LDPs), and their physical
significance has only recently been revived by works such as
Refs.~\cite{Mathisen2018, Hatano2019, Minganti2019}. Thus, since
2019 there has been a growing theoretical interest stimulated by
experimental progress~\cite{Naghiloo2019, Chen2021, Chen2022,
Zhang2022, Bu2023} in observing, understanding, and utilizing
quantum aspects of Liouvillian singularities. This includes also
closely related concepts of Liouvillian diabolical points (i.e.,
spectral degeneracies, where eigenvalues coalesce, but the
associated eigenvectors remain
orthogonal)~\cite{Minganti2021a,Minganti2021b}, hybrid LEPs (which
interpolate between HEPs and LEPs)~\cite{Minganti2020, Chen2021},
and higher-order eigenspectrum degeneracies exhibiting hybrid
properties of both diabolical and exceptional
points~\cite{Perina2022,Arkhipov2023}.

Quantum process tomography (QPT) is a procedure that enables a
complete experimental characterization of a quantum black box or,
in mathematical terms, the reconstruction of the Liouvillian
superoperator characterizing completely the dynamics of an unknown
quantum process (see reviews~\cite{NielsenBook, DAriano2003,
Mohseni2008} and references therein). Due to the formal
equivalence between processes and channels, QPT is often
considered a quantum-channel tomography. QPT was introduced in
Refs.~\cite{Chuang1997, Poyatos1997, DAriano1998} as a
generalization of quantum state tomography (QST) for
reconstructing quantum channels via reconstructing quantum output
states for various input states. Similarities between QPT and QST
include even the use of maximum-likelihood estimation to guarantee
that an experimentally reconstructed Liouvillian superoperators
(or a density matrix) really describes a physical process (or
state)~\cite{ParisBook, Fiurasek2001}). Anyway, QST and QPT are
two related but distinct procedures of quantum engineering: QST
aims to reconstruct the quantum state (density matrix) of a system
by measuring various observables. This process only tells us about
the specific state of a quantum system at a given point. While QPT
is used to fully characterize the dynamics or transformation
(quantum process) that a system undergoes. It reveals how any
input state is mapped to an output state by a quantum channel,
gate, or process. These two procedures become equivalent only in a
specific scenario when the process itself is simply an identity
operation (i.e., it leaves states unchanged), QPT effectively
reduces to QST. In this case, the only task left is to determine
the state of the system, as there is no transformation occurring.
In all other cases, QPT is more complex, as it requires
understanding the transformation effects on a full set of basis
states, while QST is limited to reconstructing the description of
just one state.

First experimental demonstrations of QPT were reported for
characterizing two-qubit gates using nuclear-magnetic-resonance
(NMR) spectroscopy~\cite{Childs2001}, and
single-~\cite{Altepeter2003, DeMartini2003} and
two-qubit~\cite{OBrien2003, Mitchell2003, OBrien2004} gates using
linear optics and conditional measurements. A multi-qubit (say
$n$-qubit) QPT can be realized by replicating ($n$ times) a given
experimental setup for a single-qubit QPT~\cite{DeMartini2003}.
This implies that the dimension of a reconstructed Liouvillian
superoperator and the complexity of QPT grows exponentially with
$n$. Recent experimental implementations of QPT (and related
tomography methods) include: trapped-ion qubit
gates~\cite{Av2020}, superconducting quantum
processors~\cite{Shukla2020, Samach2022, Pears2023}, photon
polarization damping channels~\cite{Ku2022}, and plasmonic
metamaterials operating as polarization-dependent loss channels in
quantum plasmonics~\cite{Asano2015}, etc. However, to our
knowledge, the experimental observation of LEPs of the
Liouvillians reconstructed via QPT has not been reported yet.

We analyze and experimentally implement QPT and reveal LEPs using
single-, two-, and three-qubit superconducting circuits, shown in
Fig.~\ref{fig1}, using an IBM quantum (IBMQ)
processor~\cite{IBMQ}. Note that we initially performed
simulations on the circuits, both without and with noise, as shown
in Figs.~\ref{fig2}--\ref{fig4}. Only afterward we conducted the
actual experiments.  Therefore, our main experimental results are
presented alongside the simulation results in Figs.~\ref{fig3} and
\ref{fig4}.

We argue that various experimental methods used for single-qubit
QST~\cite{Naghiloo2019} and QPT (e.g.,~\cite{Samach2022,
Howard2006}) can be modified to induce and reveal LEPs along the
lines described here. QPT can enable experimental finding not only
LEPs but also quantum diabolical points, which can reveal
dissipative phase transitions and a Liouvillian spectral
collapse~\cite{Minganti2021a, Minganti2021b}. We note that QST has
been applied across LEPs in Refs.~\cite{Chen2021, Chen2022} (see
also Ref.~\cite{Naghiloo2019}). But to our knowledge QPT has not
been applied to reveal LEPs yet. In particular, an LEP-based
quantum heat engine was studied experimentally in
Refs.~\cite{Zhang2022,Bu2023}, but neither QST nor QPT was applied
there.

Our work serves primarily as a proof-of-principle study,
demonstrating the experimental feasibility and effectiveness of
QPT for analyzing quantum system dynamics near LEPs. Another key
feature of our paper is that it is the first experimental
observation of an LEP in a system that does not exhibit any HEPs,
as highlighted in the article's title, including single-, two-,
and three-qubit systems. In our view, this result would hold
significant value on its own, even if it had been obtained without
the QPT-based approach described here, but rather through
established methods used in prior LEP-related experiments.
Additionally, we report notable physical phenomena, specifically
the direct observation of transitions between non-spiraling and
spiraling regimes. These regimes, associated with distinct decay
behaviors governed by real and complex Liouvillian eigenvalues,
were observed in physical systems (including single-, two-, and
three-qubit systems), where LEPs had not previously been
experimentally observed. Previous experimental studies were
limited to single three-level systems (qutrits).

The paper is organized as follows: In Sec.~II, we recall the LEP
formalism and describe how to detect LEPs via QPT. In Sec.~III, we
show the applicability of the method by analyzing a specific
prototype model of a lossy driven qubit exhibiting LEPs but not
HEPs. By applying completely positive maps with unitary gates, as
described in Sec.~IV, we implemented the model on IBMQ processors,
as reported in Sec.~V. The physical interpretation of transitions
observed at LEPs is explained in Sec.~VI. Section VII presents a
broader discussion of the results, including potential
generalizations of the proposed method for non-Markovian systems
and the application of complementary approaches for identifying
LEPs, followed by concluding remarks. Technical details about the
applied superoperator formalism, comparison of various equivalent
QPT methods, and our estimations of errors and measurement times
are given in Appendices.

\section{Liouvillian exceptional points and their detection}

Let us consider the dissipative evolution of a quantum system
within the Lindblad master equation. We make the standard
assumption that the system weakly interacts with a Markovian
environment. In the case of the QPT of composite systems (e.g., a
qubit and a cavity mode), it is usually also assumed that the
interaction between the subsystems (e.g., light and matter) can be
either weak or strong, but not ultrastrong, so that each of the
subsystems dissipates via a separate dissipative channel, rather
than combined channels, which would require applying a generalized
master equation~\cite{Beaudoin2011, Settineri2018, Mercurio2023}.
The expected photon output rate in the ultrastrongly coupled
light-matter systems is not directly related to the number of
photons in a cavity~\cite{Kockum2019}. Thus, a generalized QPT
should be applied which, however, is not studied here.

A general-form Lindblad master equation can be expressed via the
Liouvillian superoperators $\LL$~\cite{BreuerBook, HarocheBook}
($\hbar=1$) as
\begin{equation}\label{Eq:ME}
\frac{\partial}{\partial t} {\hat\rho} = \LL \rhot =- i [\hat{H},
\rhot] + \sum_{\mu} \DD[\hat{\Gamma}_{\mu}] \rhot,
\end{equation}
acting on the density matrix $\rhot$ of the system described by a
Hermitian Hamiltonian $\hat{H}$ at an evolution moment $t$. For
the clarity of our presentation, the standard matrix
representation of superoperators is recalled in Appendix~A. The
Lindbladian dissipators $\DD [\hat{\Gamma}_\mu]$ are given by
\begin{equation}
  \DD [\hat{\Gamma}_\mu] \rhot = \hat{\Gamma}_\mu \rhot
\hat{\Gamma}_\mu^\dagger - \frac{1}{2}[\hat{\Gamma}_\mu^\dagger
    \hat{\Gamma}_\mu \rhot + \rhot \hat{\Gamma}_\mu^\dagger
    \hat{\Gamma}_\mu],
  \label{dissipators}
\end{equation}
where $\hat{\Gamma}_\mu$ are quantum jump operators with a clear
interpretation in the quantum-trajectory approach (also referred
to as the wave-function Monte Carlo method)~\cite{Dalibard1992,
Carmichael1993, Molmer1993, Plenio1998, Daley2014}. Consequently,
one can also introduce an effective NHH,
\begin{equation}
  \hat{H}_{\rm eff} = \hat{H} - \frac{i}{2} \sum
\hat{\Gamma}_\mu^\dagger \hat{\Gamma}_\mu,
  \label{H_eff}
\end{equation}
and rewrite Eq.~(\ref{Eq:ME}) as
\begin{equation}\label{Eq:Lindblad2}
 \LL \rhot =- i \left[\hat{H}_{\rm
eff}\hat{\rho}(t) - \hat{\rho}(t) \hat{H}_{\rm eff}^\dagger
\right]+ \sum_\mu \hat{\Gamma}_\mu \rhot \hat{\Gamma}_\mu^\dagger.
\end{equation}
This master equation encompasses the terms describing a continuous
non-unitary dissipative evolution via $\hat{H}_{\rm eff}$, and the
quantum-jump term. A quantum jump is a sudden stochastic change of
the wave-function corresponding to the loss or gain of a system
excitation due to the interaction with the environment, which
monitors (``measures'') the system~\cite{CarmichaelBook,
HarocheBook}. This quantum-trajectory interpretation of the master
equation is physically very intuitive and reveals the importance
of effective NHHs, which are used in standard quantum mechanics
and are not limited to \PT-symmetric systems. They describe
continuous losses of energy, coherence, and quantum information of
a system into its environment. Moreover, this master equation
interpretation also reveals crucial role of quantum jumps. Their
omission can be justified in the semiclassical limit or by
postselecting quantum trajectories.

We consider the eigenproblems:
\begin{eqnarray}
  \hat{H}_{\rm eff} \ket{E_n} &=& E_n \ket{E_n}, \label{eigenproblem1} \\
  \LL\hat\rho_n&=&\lambda_n\hat\rho_n, \label{eigenproblem2} \\
  \LL^\dagger\hat\sigma_n&=&\lambda^*_n\hat\sigma_n,\label{eigenproblem3}
\end{eqnarray}
where $E_n$ and $\ket{E_n}$ are the eigenvalues and eigenvectors
of the NHH operator; while $\lambda_n$, $\hat\rho_n$, and
$\hat\sigma_n$ are the eigenvalues, as well as the right and left
eigenmatrices of the Liouvillian superoperator, respectively. With
these eigenspectra, HEPs and LEPs can be found. Note that
$\hat\rho_n$ and $\hat\sigma_n$ for a given $n$ are mutually
orthogonal. However, different $\hat\rho_n$ (as well as
$\hat\sigma_n$) are not, in general, orthogonal. The real parts of
$\lambda_n$ for any $n$ is non-positive and describes a relaxation
rate towards the system's steady state~\cite{BreuerBook}. By
representing the eigenmatrices $\hat\rho_n$ and $\hat\sigma_n$ as
vectors $\ket{\tilde\rho_n}$ and $\bra{\tilde\sigma_n}$,
respectively, and treating the Liouvillian superoperator ${\cal
L}$ as a matrix $\tilde {\cal L}$, Eqs.~(\ref{eigenproblem2}) and
(\ref{eigenproblem3}) can be rewritten, respectively, as
\begin{equation}
  \tilde {\cal L} \ket{\tilde\rho_n} =
\lambda_n\ket{\tilde\rho_n}\quad {\rm and} \quad
\bra{\tilde\sigma_n}\tilde {\cal L} =
\lambda_n\bra{\tilde\sigma_n}.
  \label{spectrumLtilde}
\end{equation}

The LEPs of $\tilde {\cal L}$ can be calculated by applying the
standard superoperator formalism~\cite{Minganti2019}. Such LEPs
can be found experimentally via the QPT based on $6\times 6$
projectors, i.e., assuming that the input and output states (or
projections) are the eigenstates of all the Pauli operators:
$|\In_i\rangle,|\Out_j\rangle \in
  \{|x_+\rangle,|x_-\rangle,|y_+\rangle,|y_-\rangle,|z_+\rangle,|z_-\rangle\},$
where $|x_{\pm}\rangle = \frac{1}{\sqrt{2}}\left(|0\rangle
\pm|1\rangle\right)$, $|y_{\pm}\rangle = \frac{1}{\sqrt{2}}
\left(|0\rangle \mp i|1\rangle\right)$,
$|z_+\rangle\equiv|0\rangle$, and $|z_-\rangle\equiv|1\rangle$.
These projections can be used for the QPT of a transmon qubit,
where $|0\rangle$ ($|1\rangle$) corresponds to its ground
(excited) state. Thus, for a dissipative and/or amplified process
described by the master equation with a Liouvillian $\LL$, one can
measure all its elements
$  L'_{ij}=\langle \Out_j|\LL\big(\hat\rho=|\In_i\rangle\langle
   \In_i|\big)|\Out_j\rangle,$
and, thus, reconstruct the full $6\times 6$ matrix $L=[L'_{ij}]$,
which represents $\LL$. We refer to this approach as Method 1.
Other approaches can also be applied, including Methods 2 and 3
described in Appendix~B. All these methods reveal the same LEPs
under perfect measurement conditions as shown in detail in
supplementary materials in Supplement~1 and briefly explained in
Appendix~C.

The dynamics of an open quantum system is governed by
Eq.~(\ref{Eq:ME}). For short evolution steps $dt$, this
corresponds to
\begin{equation}
  \hat\rho(t+dt) = (\LL dt + 1) \hat\rho(t) \equiv S\hat\rho(t).
  \label{S}
\end{equation}
This short-time evolution of a quantum state $\hat\rho(t)$ under
the non-Hermitian dynamics, where $S$ is the effective quantum
operation, is the subject of QPT. Note that $S$ has the same
spectral decomposition as $\LL$ up to an affine transformation for
all eigenvalues related to scaling by $dt$ and shifting by $1$.
Thus, we can study LEPs by performing QPT on $S$. We choose $dt$
depending on the specific form of $\LL$, being small enough to
realize specific dynamics. If the operation $S$ is applied to a
system $n$ times, the evolution is effectively described by the
master equation for the evolution time $ndt.$

\section{A lossy driven qubit} \label{}

In our experiment performed on an IBMQ processor~\cite{IBMQ}, we
applied QPT to reveal LEPs in a driven lossy single-qubit
(spin-1/2) prototype model, which exhibits LEPs but not
HEPs~\cite{Minganti2019, Minganti2020}. Specifically, the system
is described by the Hamiltonian $\hat{H} = (\omega/2)
\hat{\sigma}_z,$ and decays through three competing channels
($\hat\sigma_x$, $\hat\sigma_y$, and $\hat\sigma_-$), as described
by the Liouvillian,
\begin{equation}
    \LL \rhot = - i [\hat{H}, \rhot] +
    \Big(\gamma_- \DD[\hat{\sigma}_-]
     +\gamma_x\mathcal{D}[\hat\sigma_x]
    +\gamma_y\mathcal{D}[\hat\sigma_y]\Big)\rhot,
\label{QubitLiouvillian}
\end{equation}
where $\hat\sigma_{x,\, y, \, z}$ are the Pauli matrices, and
$\hat{\sigma}_{\pm} = (\hat\sigma_{x} \mp i \hat\sigma_{y})/2$ are
the qubit raising and lowering operators, respectively. We note
that some typos in the corresponding equation in
Ref.~\cite{Minganti2019} have been corrected here to ensure that
the numerical results can be accurately reproduced. The terms of
the master equation describe, respectively: (1) oscillations, (2)
erroneous bit flips at a rate $\gamma_-$, and (3,4) dissipation
with rates $\gamma_x$ and $\gamma_y$ along the $x$- and $y$-axes
of the Bloch sphere, respectively. This dissipative model exhibits
a $\cal{Z}_2$ symmetry, as it remains invariant under the
transformation $\hat{\sigma}_- \rightarrow -\hat{\sigma}_-$
\cite{Albert2014, Minganti2018}. It is a prototype model that can
be applied to various systems beyond the one studied here. For
instance, it can describe a spin-$\frac{1}{2}$ particle in a
uniform magnetic field along the $z$-axis, assuming relaxation of
the particle along the $x$- and $y$-axes, and allowing for
spin-flip errors. The oscillations induced by the Hamiltonian, the
dissipation occurring along the $x$ and $y$ axes, and the spin
flipping induced by $\hat{\sigma}_-$ determine how quickly the
system reaches a steady state.

The lack of HEPs is evident due to the diagonal structure of the
effective NHH in the standard computational basis, i.e.,
\begin{equation}
    \hat{H}_{\rm eff}=\frac{1}{2} \diag([ \omega -i \gamma_{x}- i
    \gamma_{y} -i \gamma_-,-\omega -i \gamma_{x} - i\gamma_{y}]),
\label{eq:effHamiltonian}
\end{equation}
as there is no way to adjust the parameters to make the two
eigenvalues equal. Despite this, the Liouvillian still exhibits
LEPs. Specifically, one finds the eigenvalues~\cite{Minganti2019}:
\begin{eqnarray}
  \lambda_0 &=& 0, \nonumber\\
  \lambda_{1,2} &=& -\frac{\gamma_-}{2}-\gamma _x
  -\gamma_y\pm\Omega, \nonumber\\
  \lambda_3&=&\gamma _--2 \left(\gamma_y+\gamma_x\right),
\label{lambda_i}
\end{eqnarray}
together with the corresponding right eigenmatrices:
\begin{eqnarray}
\eig{0} &\propto&\diag([\gamma _x+\gamma _y, \gamma _x+\gamma _y
+\gamma _- ]),\nonumber\\
\eig{1, \, 2}&\propto&
\mathrm{antidiag}([-i \omega \pm \Omega,\gamma _x-\gamma_y]),\nonumber\\
\eig{3}&\propto& \diag([-1,1]), \label{rho_i}
\end{eqnarray}
where $\Omega^2=\gamma_x^2+\gamma_y^2-2 \gamma_x \gamma_y -\omega
^2$. See Appendix~F for more analytical results. In the case
$\gamma_{y}>\omega$, this Liouvillian exhibits two LEPs at
$\gamma^{\pm}_{x}\equiv\gamma_{y}\pm\omega$. We study this
configuration experimentally by setting $\gamma_-=0$ and
$\gamma_y=2\omega$. Figures~\ref{fig2}-\ref{fig4} show theoretical
eigenvalues $\lambda_{0,1}$ and modified eigenvalues, which are
obtained via a more-realistic QPT simulation assuming white noise.
Specifically, experimental pure-like states, which are the input
states for QPT, are always mixed with some amount of white noise.
This undesired effect was included in our refined simulations.
These simulated eigenvalues are compared with $\lambda^{\rm
exp}_n$ reconstructed from our single-, two-, and three-qubit
experiments.

\begin{figure}[htp!]
(a)\\
   \includegraphics[width=\linewidth]{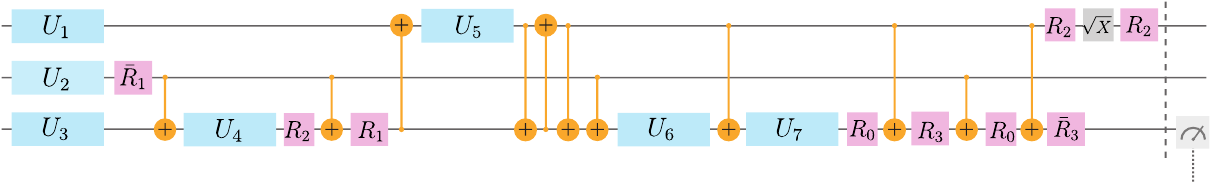}
(b)\\ \vspace*{3mm}
   \includegraphics[width=\linewidth]{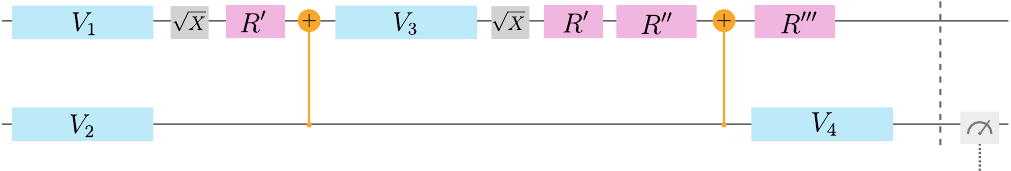}
(c)\\ \vspace*{3mm}
   \includegraphics[width=0.8\linewidth]{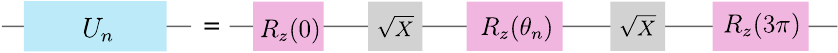}
\\(d)\\ \vspace*{3mm}
   \includegraphics[width=0.55\linewidth]{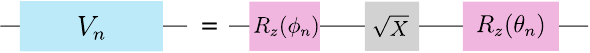}
\caption{(a) Three-qubit and (b) two-qubit circuits optimized for
the Nairobi quantum processor by Qiskit and applied in our
experiments. Gates $U_n\equiv U(\theta_n)$ and $V_n\equiv
V(\phi_n,\theta_n)$ are implemented by the sequences of basic
gates shown in (c) and (d), respectively. We set the following
phases for $U_n$, $V_n$, $R_n\equiv R_z(\zeta_n)$, and $\bar
R_n\equiv R_z(-\zeta_n)$: $U_1=U(5.7153)$, $U_2=U(4.7788)$,
$U_3=U(5.6341)$, $U_4=U(2.7850)$, $U_5=U(3.3259)$,
$U_6=U(3.0600)$, and $U_7=U(3.4146)$; $V_1=V(0,2\pi)$,
$V_2=V(-\pi/2,0.0332)$, $V_3=V(0,\pi)$, and $V_4=V(\pi/2,\pi/2)$;
$R_0=R_z(0)$, $R_1=R_z(0.0460)$, $R_2=R_z(\pi/2)$,
$R_3=R_z(1.5248)$, $R'=R_z(3\pi)$, $R''=R_z(-1.5375)$, and
$R'''=R_z(1.6040)$.} \label{fig1}
\end{figure}

\begin{figure}[htp!]
\flushleft \hspace{1.5cm} (a) \hspace{4.5cm} (b) \hspace{4.5cm}
(c) \centerline{
\includegraphics[height=4.3cm]{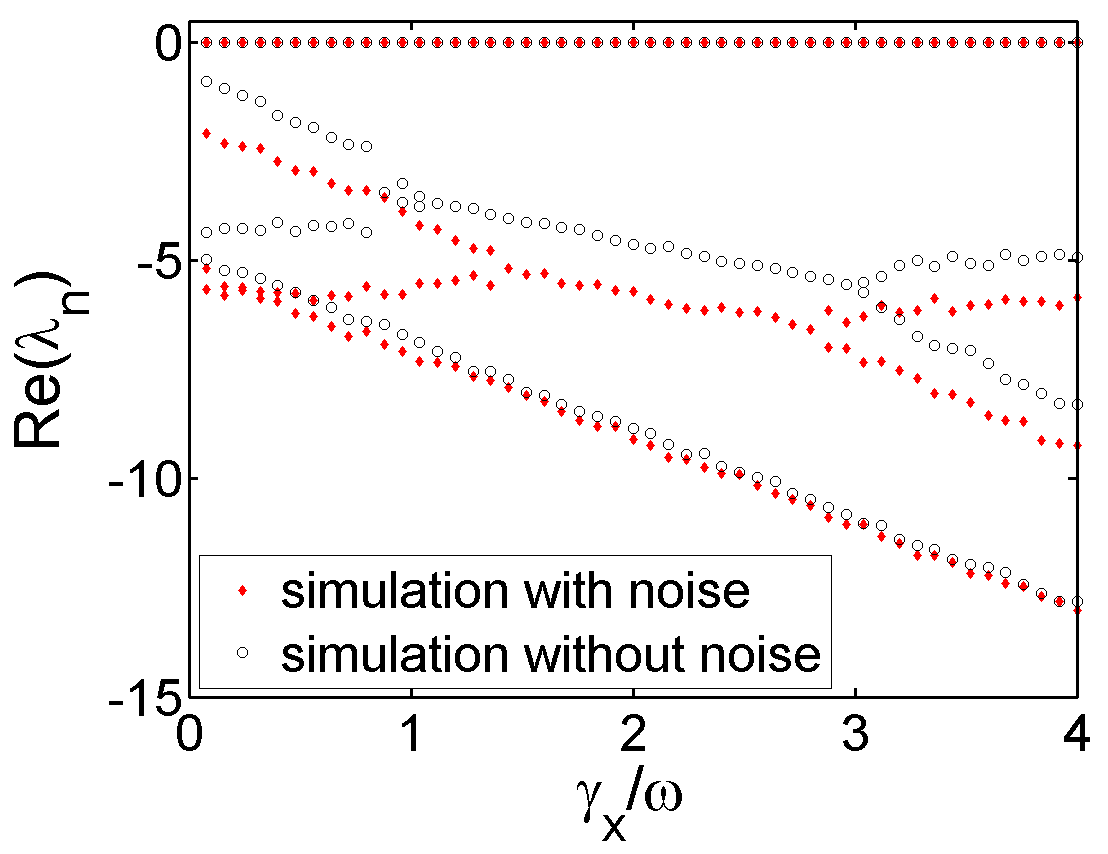}
\includegraphics[height=4.3cm]{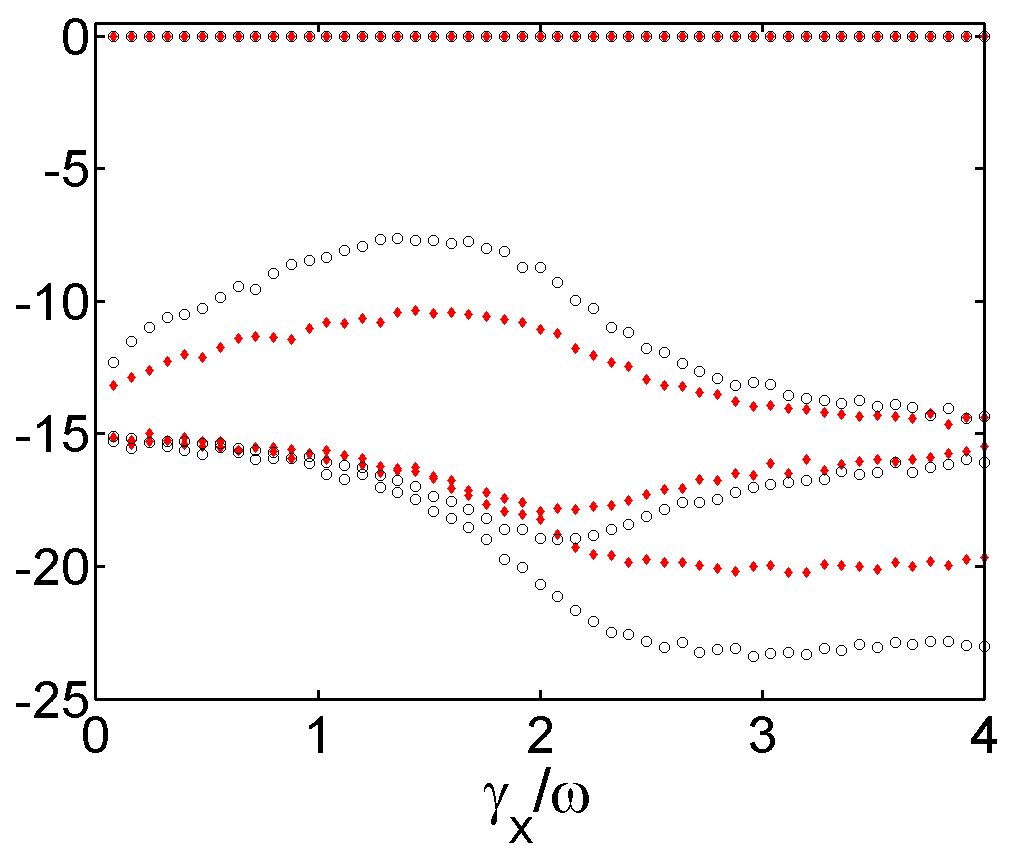}
\includegraphics[height=4.3cm]{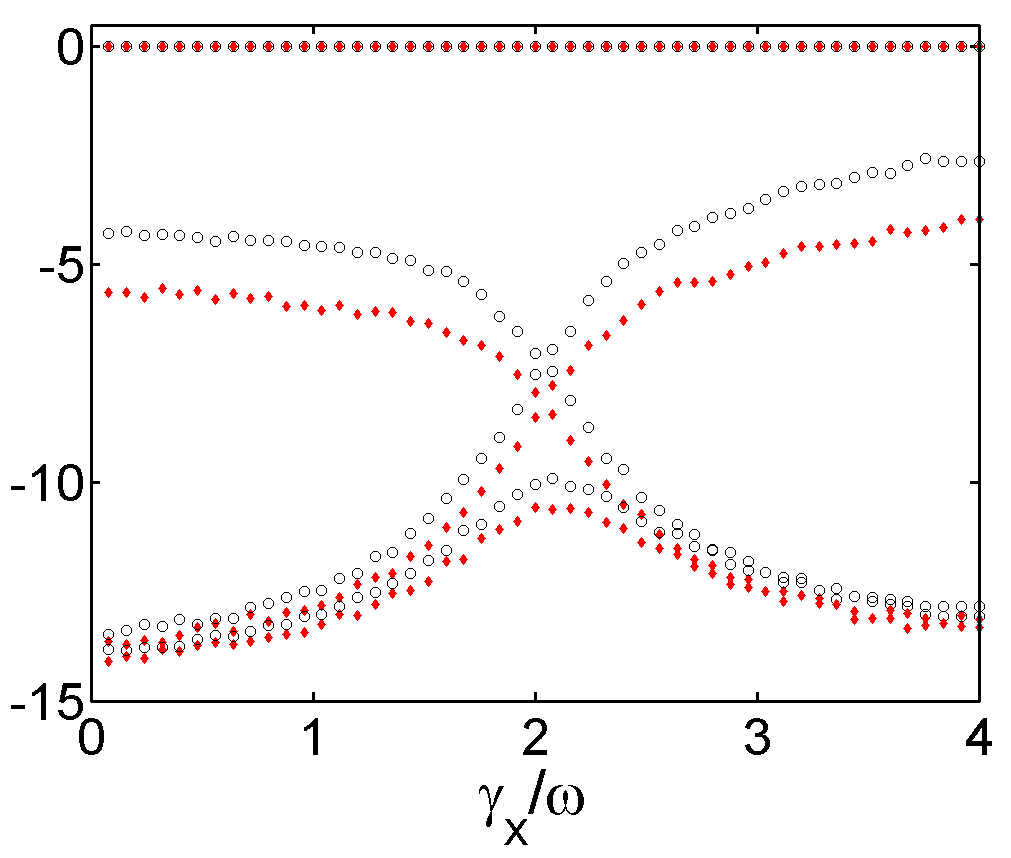}}
\caption{Real parts of the eigenvalues, Re$(\lambda_n)$, of the
Liouvillians simulated for the three-qubit circuit depicted in
Fig.~\ref{fig1}(a), with measurements conducted on a single qubit:
(a) third, (b) second, and (c) first qubit. Simulations were
carried out both with (red diamonds) and without (black circles)
the inclusion of experimental noise.}\label{fig2}
\end{figure}

\begin{figure}[htp!]
\flushleft \hspace{3.5cm} (a) \hspace{6cm} (b) \hspace{4.5cm}
\centerline{
\includegraphics[height=5cm]{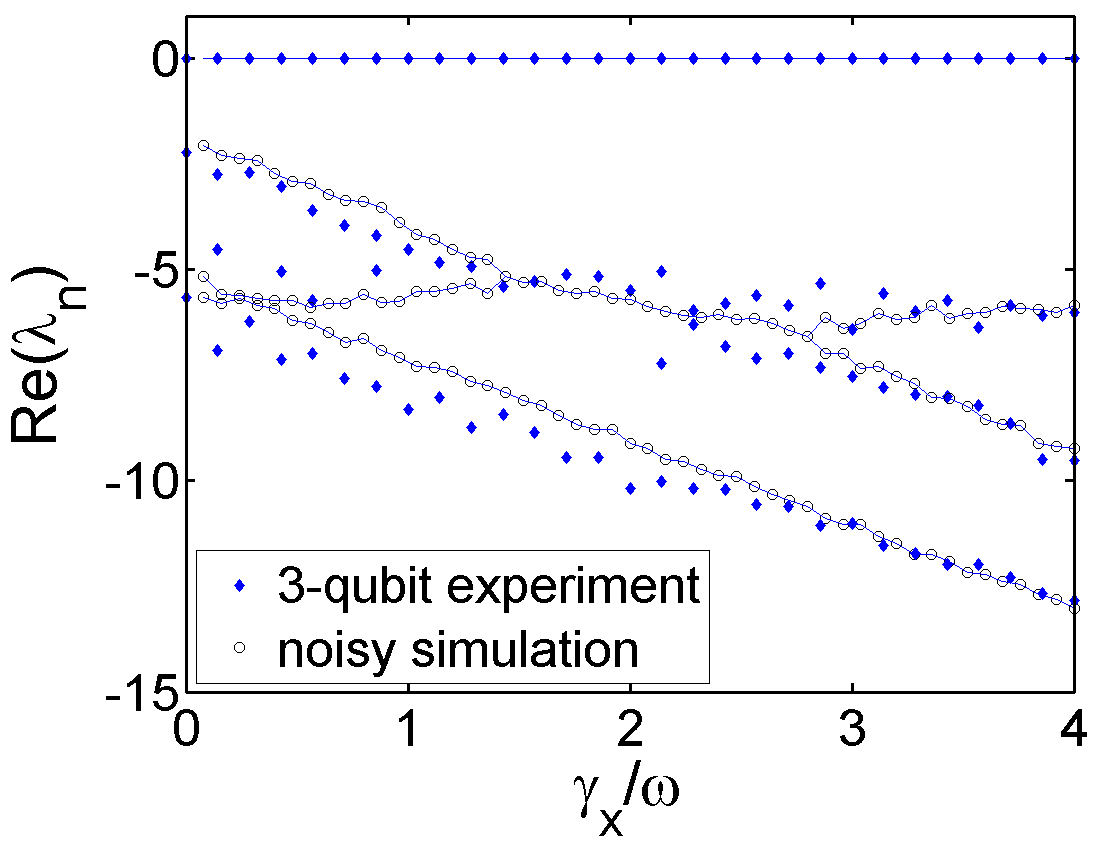}
\includegraphics[height=5cm]{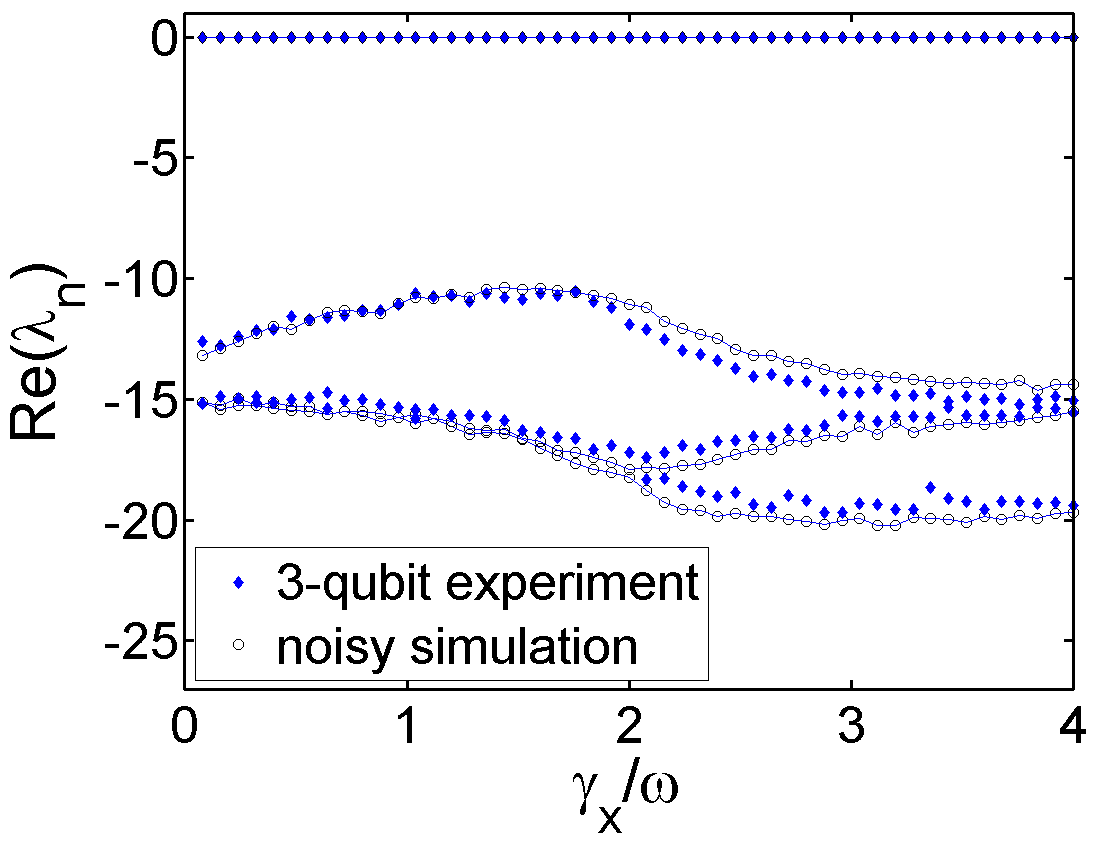}}
\caption{Experimental and simulated real parts of eigenvalues
obtained using the 3-qubit circuit, with measurements performed on
(a) the third qubit and (b) the second qubit. We note that
achieving a close match between the simulated and experimental
curves is highly sensitive to the noise mitigation techniques
applied in these calculations.} \label{fig3}
\end{figure}
\begin{figure}[htp!]
   \includegraphics[width=\linewidth]{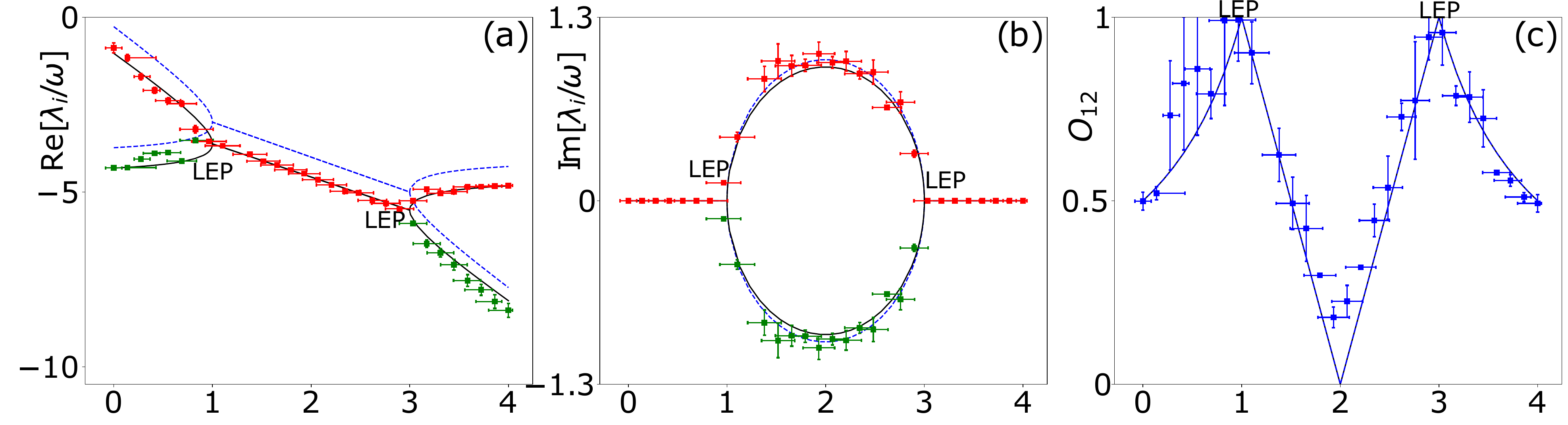}

   \includegraphics[width=\linewidth]{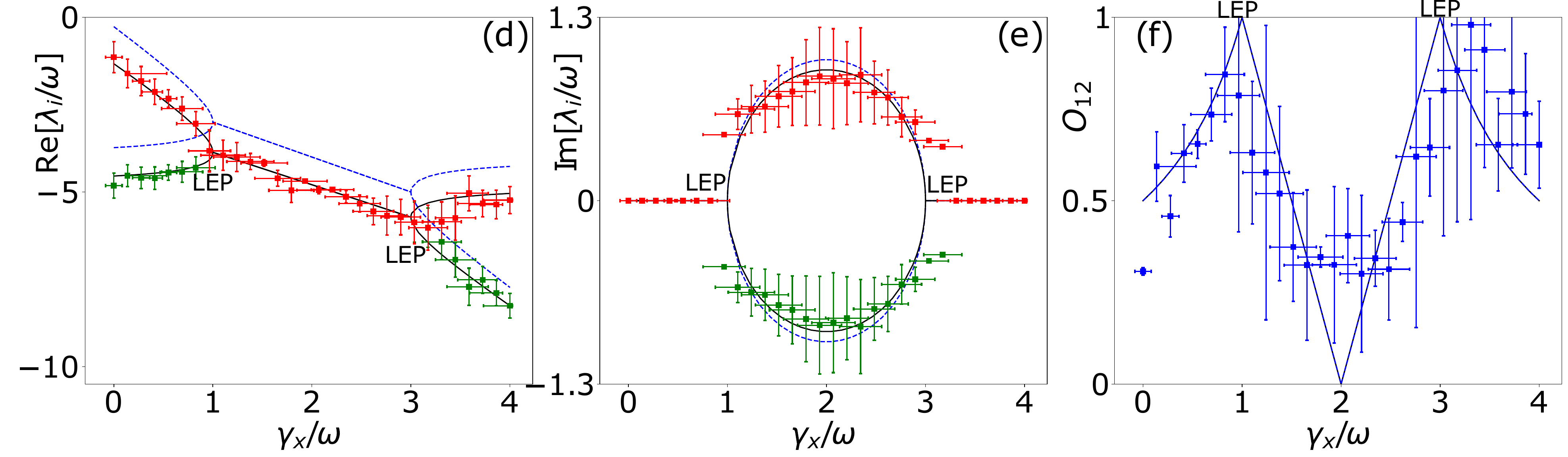}
\caption{\label{fig4} Real (a,d) and imaginary (b,e) parts of the
eigenvalues $\lambda_1$ and $\lambda_2$ of the experimental and
theoretical Liouvillians, and the overlaps (c,f),
$O_{12}=|\langle\tilde\sigma_{1}|\tilde\rho_{2}\rangle|$, of the
eigenmatrices $\langle \tilde\sigma_{1}|$ and
$|\tilde\rho_{2}\rangle$. In the experimental data, $\lambda_1$ is
indicated by red squares, and $\lambda_2$ by green squares when
$\lambda_1 \neq \lambda_2$. Experimental results are reconstructed
from single-qubit (a,b,c) and two-qubit (d,e,f) measurements
performed on an IBMQ processor~\cite{IBMQ} (squares) and shown
together with the corresponding theoretical predictions including
white noise (black solid curves) and without it (blue broken
curves). Each measurement was carried out with 20,000 shots and
$\omega dt =1/15$. To enhance plot clarity, the two less relevant
real eigenvalues (specifically, $\lambda_{0}=0$ and the smallest
eigenvalue $\lambda_{3}$) are omitted here but are displayed in
other figures. The results shown in panels (a,b,c) for the
single-qubit circuit are clearly less noisy than those in panels
(d,e,f) for the two-qubit circuit and are significantly less noisy
than the results in Fig.~\ref{fig3}(a), which were obtained using
the three-qubit circuit. The selected eigenvalues from panels (a,
b) and (d, e) are listed explicitly in Tables~1 and 2,
respectively.}
\end{figure}
\begin{table}[ht!]
\centering
\begin{tabular}{ccccccccc}
 \hline \hline
$m$ & $\gamma_x/\omega$ &
 Re$\lambda_1$ & Im$\lambda_1$ &
 Re$\lambda_2$ & Im$\lambda_2$ &
 $\lambda_3$ & Gap & Remarks\\
 \hline

1 & 0    &  -0.88 & 0    & -4.31 & 0     & -4.95  & 0.88 & Smallest spectral gap \\
2 & 0.83 &  -3.21 & 0    & -3.52 & 0     & -6.60  & 3.21 & Non-spiraling point nearest LEP 1\\
3 & 0.97 &  -3.55 & 0.13 & -3.55 & -0.13 & -6.89  & 3.55 & Spiraling point closest to LEP 1  \\
4 & 2.07 &  -4.64 & 1.04 & -4.64 & -1.04 & -9.10  & 4.64 & Largest spiraling \\
5 & 2.90 &  -5.48 & 0.33 & -5.48 & -0.33 & -10.77 & 5.48 & Spiraling point closest to LEP 2 \\
  &      &        &      &       &       &        &      & and largest spectral gap \\
6 & 3.03 &  -5.25 & 0    & -5.90 & 0     & -11.01 & 5.25 & Non-spiraling point nearest LEP 2\\
7 & 4.00 &  -4.82 & 0    & -8.38 & 0     & -12.92 & 4.82 & $\lambda_3^{(7)}=\min_{n,m} \lambda_n^{(m)}$  \\
\hline \hline
\end{tabular}
\caption{Experimental nonzero eigenvalues $\lambda_n \equiv
\lambda_n^{(m)}$ selected from Figs.~\ref{fig4}(a)
and~\ref{fig4}(b) obtained using the single-qubit circuit: Real
and imaginary parts of $\lambda_n$ for chosen values of the
damping coefficient $\gamma_x$. Both $\lambda_n$ and $\gamma_x$
are given in units of $\omega$. The error bars are indicated in
the figures.  The Liouvillian spectral gap (referred to here
simply as the ``gap'') is defined as the non-zero eigenvalue with
the smallest modulus of its real part. The minimal measured
eigenvalue is given by $\lambda_{\min} = \min_{n,m}
\lambda_n^{(m)}=\lambda_3^{(7)}$ among all eigenvalues $n = 1, 2,
3$ for the $M=29$ measured Liouvillians ($m = 1, \ldots, M$)
plotted in the figures. The trivial zero eigenvalue,
$\lambda_0^{(m)}=0$, is omitted. The terms `closest,' `largest,'
and `smallest' specifically refer to the extremal eigenvalues of
these experimental Liouvillians.} \label{table1}
\end{table}

\begin{table}[ht!]
\centering
\begin{tabular}{cccccccccc}
 \hline \hline
$m$ & $\gamma_x/\omega$ &
 Re$\lambda_1$ & Im$\lambda_1$ &
 Re$\lambda_2$ & Im$\lambda_2$ &
 $\lambda_3$ & Gap & Remarks\\
 \hline
1 & 0    &  -1.14 & 0    & -4.82 & 0     & -5.24  & 1.14 & Smallest spectral gap \\
2 & 0.83 &  -3.04 & 0    & -4.31 & 0     & -6.67  & 3.04 & Non-spiraling point nearest LEP 1\\
3 & 0.97 &  -3.82 & 0.47 & -3.82 & -0.47 & -7.11  & 3.82 & Spiraling point closest to LEP 1 \\
4 & 2.48 &  -5.14 & 0.89 & -5.14 & -0.89 & -9.66  & 5.14 & Largest spiraling \\
5 & 3.31 &  -6.02 & 0.38 & -6.02 & -0.38 & -11.40 & 6.02 & Spiraling point closest to LEP 2 \\
  &      &        &      &       &       &        &      & and largest spectral gap \\
6 & 3.45 &  -5.85 & 0    & -6.43 & 0     & -11.64 & 5.85 & Non-spiraling point nearest LEP 2 \\
7 & 0    &  -5.23 & 0    & -8.26 & 0     & -12.73 & 5.23 & $\lambda_3^{(7)}=\min_{n,m} \lambda_n^{(m)}$  \\
\hline \hline
\end{tabular}
\caption{Similar to Table~\ref{table1}, but this table presents
the experimental eigenvalues selected from Figs.~\ref{fig4}(d)
and~\ref{fig4}(e), which were obtained from $M=30$ Liouvillians
measured on the two-qubit circuit, shown in Fig.~\ref{fig1}(b).}
\label{table2}
\end{table}

\section{Implementing completely positive  maps with unitary
gates}

To implement non-Hermitian dynamics using only unitary operations,
we purify (coherify) the quantum process by embedding it in a
larger Hilbert space, where the joint evolution of the system and
its environment is unitary. In particular, starting with the
superoperator $S,$ we find its Choi representation $\hat\chi$.
Depending on the number of nonzero eigenvalues of the Choi matrix,
we choose the dimension of the required ancillary system. This
approach to implement completely positive maps is well known (see,
e.g., Ref.~\cite{Fiurasek2001PRA}). As we can implement an
arbitrary unitary operation on a programmable quantum computer, we
use this approach to demonstrate LEPs experimentally with a noisy
intermediate-scale quantum (NISQ) processor.

Completely positive (CP) maps are linear maps that preserve the
positivity of density matrices. To express a CP map
$\mathcal{E}_{\mathcal{H}}$ between the Hilbert spaces
$\mathcal{H}$ and $\mathcal{K}$ as a unitary operator we can use
the Choi-Jamio\l{}kowski
isomorphism~\cite{Jamiolkowski1972,Choi1975} between the map and
operator $\hat\chi$. The associated quantum operation can be
expressed as
\begin{equation}
  \hat{\rho}_{\mathrm{out}}=
\mathrm{tr}_{\mathcal{H}}\left[\hat\chi
\hat{\rho}_{\mathrm{in}}^{\mathrm{T}}\otimes
\hat{\mathds{1}}_{\mathcal{K}} \right],
  \label{Choi}
\end{equation}
where the operator, which is isomorphic to the map, reads
\begin{equation}
  \hat\chi=\mathcal{E}_{\mathcal{H}}\otimes\mathcal{I}_{\mathcal{H}}(\ket{\phi}\bra{\phi}),
  \label{chi}
\end{equation}
where
$\mathrm{tr}_{\mathcal{K}}[\hat\chi]=\hat\openone_{\mathcal{H}},$
 $\ket{\phi}=\sum_{j=1}^{\mathrm{dim}\mathcal{H}}
\ket{j}_1\ket{j}_2$,  $\mathcal{I}$ is an identity map, and
$\hat\openone_{\mathcal{H}}$ denotes the identity operator on
$\mathcal{H}$. The corresponding Kraus decomposition reads
\cite{Fiurasek2001PRA}:
\begin{equation}
  \hat{\rho}_{\mathrm{out}}=\mathcal{E}(\hat{\rho}_{\mathrm{in}})=
\sum_l \hat{A}_l \hat{\rho}_{\mathrm{in}} \hat{A}_l^\dagger,
  \label{Kraus}
\end{equation}
where
 $\sum_l \hat{A}_l^\dagger \hat{A}_l = \hat{\mathds{1}}_{\mathcal{H}}.$
which can be rewritten, by substituting $A_{ki}^{(l)}\equiv
\bra{k}\hat{A}_l\ket{i}$, as
 $\sum_{k,l}A_{ki}^{*(l)}A_{kj}^{(l)}=\delta_{ij}$.
The number of the $\hat{A}_l$ operators corresponds to the number
of nonzero eigenvalues of the $\hat{\chi}$ matrix; $\hat{\chi}$
and $\hat{A}_l$ can be related to each other by the eigenvalues
$r_l$ and the eigenstates $\ket{\pi_l}$ of the $\hat{\chi}$
operator: $A_{ki}^{(l)} =
\sqrt{r_l}\bra{k}\langle{i}|{\pi_l}\rangle,$ where $\ket{i}\in
\mathcal{H}$ and $\ket{k}\in \mathcal{K}$ are the states in the
input and output Hilbert spaces, respectively. Finally, we have
\begin{equation}\label{eq:3Qexp}
\hat{\rho}_{\mathrm{out}}= \mathrm{tr}_{\mathrm{env}}[\hat{U}
\hat{\rho}_{\mathrm{in}}\otimes (\ket{0}\bra{0})_{\mathrm{env}}
\hat{U}^\dagger],
\end{equation}
where
$\hat{U}=\sum_l\hat{A}_l\otimes(\ket{l}\bra{0})_{\mathrm{env}}$ is
the unitary operation decomposable into quantum gates.

For the discussed driven lossy qubit model, the number of the
nonzero eigenvalues of the Choi matrix is $\leq 4$. The simplest
single-qubit circuit implementing the CP map applies a unitary
operation corresponding to $A^{(l)}$ at random with probability
$r_l$, as described by the Kraus representation. When applied
repeatedly to the initial quantum state, the resulting final state
approximates the time-evolved quantum state of the simulated
system. However, this is not a fully quantum simulation of the
quantum dynamics, as an external random number generator is
required.

The second simplest experiment is embedded in a two-qubit Hilbert
space and utilizes two-qubit unitary operations and a single-qubit
environment, $\hat{\rho}_{\mathrm{env}}^{(1)}$, and reads as
\begin{equation}\label{eq:2Qex}
\hat{\rho}_{\mathrm{out}}\otimes \hat{\rho}_{\mathrm{env}}^{(1)}
    =\sum_{m=0,2}\hat{\cal A}_{m}
    \left(\hat{\rho}_{\mathrm{in}}
    \otimes \ket{0}\bra{0}\right)\hat{\cal A}^\dagger_{m},
\end{equation}
where
\begin{equation}
  \hat{\cal A}_{m}=\sum_{l=0,1}\hat{A}_{l+m} \otimes
\ket{l}\bra{0},
  \label{A_m}
\end{equation}
which requires using two random two-qubit operations ($m=0,2$).
Finally, a completely coherent quantum three-qubit experiment
requires applying a single unitary operation and a two-qubit
environment, as described by Eq.~(\ref{eq:3Qexp}).

When working with a programmable quantum computer, we are mostly
limited to applying noisy unitary operations and imperfect
readout. There are many approaches towards implementing qubits on
quantum computers. Here we focus on transmon qubits, which are
nowadays commonly used in superconducting quantum processors.
These processors are able to implement sets of elementary
instructions containing both single- and two-qubit unitary
operations. Not every two qubits in a quantum chip are coupled
directly. This requires transpiling a given unitary operation into
elementary gates according to a coupling map of a given quantum
processor. While the fidelities of single- and two-qubit
operations are typically high, the gate errors can accumulate to
an unacceptable level. If the time required to execute all the
gates is comparable to  the coherence time $T_2$ of the used
transmons, then the results are largely affected by decoherence.
All of these limitations should be taken into account when
designing an experiment. It is evident that the experimentally
reconstructed dynamics is usually perturbed with respect to the
expected one. The effective perturbations in the eigenvalues
($\delta \lambda$) and eigenmatrices ($|\delta
\tilde\rho_{n}\rangle$ and $\langle\delta\tilde\sigma_{n}|$) of an
experimental Liouvillian $L^{\exp}=L_0+\delta L$, with the
eigenspectrum obtained experimentally, compared to the ideal
unperturbed Liouvillian $L_0$ and its eigenspectrum [denoted with
superscript ${(0)}$], can be estimated as~\cite{James2001} (see
Appendix~D for details):
\begin{eqnarray}
\delta \lambda_n &\approx& \langle\tilde\sigma_{n}^{(0)}|\delta{L}
|\tilde\rho_{n}^{(0)}\rangle,
 \label{delta_lambda} \\
|\delta \tilde\rho_{n}\rangle &\approx& -  \sum_{i\;(i\neq n)}
\left(\frac{ \langle\tilde\sigma^{(0)}_{i}| \delta{L}
|\tilde\rho^{(0)}_{n}\rangle}{\lambda^{(0)}_{i}-\lambda^{(0)}_{n}}\right)
|\tilde\rho^{(0)}_{i}\rangle,
 \label{delta_rho} \\
\langle\delta\tilde\sigma_{n}| &\approx& - \sum_{i\;(i\neq n)}
\left(\frac{\langle\tilde\sigma^{(0)}_{n}| \delta{L}
|\tilde\rho_{i}^{(0)}\rangle}{\lambda^{(0)}_{i}-\lambda^{(0)}_{n}}\right)
\langle\tilde\sigma^{(0)}_{i}|.
 \label{delta_sigma}
\end{eqnarray}
The $y$-axis error bars depicted in Fig.~\ref{fig4} were
calculated according to these equations. Thus, based on these
estimations, we can select the most noise-robust experimental
strategy. Moreover, the uncertainties of $\gamma_x$, as plotted in
Fig.~\ref{fig4}, are estimated in Appendix~E.

\section{Experiment}

Here, we explain how we conducted experiments using the single-,
two-, and three-qubit circuits, as illustrated in Fig.~\ref{fig1}
for the latter two cases, to demonstrate the LEPs in the model
described by Eq.~(\ref{QubitLiouvillian}).

The model we experimentally implemented is, in fact, more general
than that given by Eq.~(\ref{QubitLiouvillian}), as we constructed
a three-qubit model, which is then reduced to three different
single-qubit models. Specifically, by measuring one of the three
qubits and tracing out the other two, we could derive the
previously studied model, along with two additional models that
have not been examined before.

Figure~\ref{fig2} illustrates the real parts of all the
eigenvalues of the Liouvillians that describe the dynamics of the
three single-qubit dissipative models based on our simulated
experiments using the three-qubit circuit shown in
Fig.~\ref{fig1}(a). Furthermore, we conducted experiments on this
circuit, measuring the third qubit to obtain the eigenvalues
displayed in Fig.~\ref{fig3}(a) and the second qubit for those in
Fig.~\ref{fig3}(b). These figures present our experimental results
alongside their simulations from Fig.~\ref{fig2} for both models.

For the sake of simplicity and clarity, we hereafter focus on the
specific cases illustrated in Figs.~\ref{fig2}(a), \ref{fig3}(a),
and \ref{fig4}, which correspond to the model given by
Eq.~(\ref{QubitLiouvillian}).

In our experiments the system
$\hat{\rho}_{\mathrm{in}}\equiv\rhot$ was prepared in one of the
six input states, which are the eigenstates of the three Pauli
operators. Then, the evolution under a given map was applied.
Finally, we measure
$\hat{\rho}_{\mathrm{out}}\equiv\hat{\rho}(t+dt)$ in the $x$, $y$,
and $z$ bases to reconstruct $L$ (and, consequently, $\LL$). The
results of the experiments, conducted on an IBMQ processor (i.e.,
Nairobi)~\cite{IBMQ}, were plotted in Figs.~\ref{fig3} and
\ref{fig4}. The measurements were performed for $30$ points with
20,000 shots per experiment, and the evolution step was $\omega
dt=1/15$. Quantum processors are error-sensitive due to their
susceptibility to noise and decoherence. To mitigate errors, we
used dynamic decoupling. This involves applying a sequence of
pulses to each qubit to protect it from ambient noise. The idea
behind the method is to repeatedly apply a series of the inversion
or refocussing pulses that reverse the effect of noise. These
pulses effectively separate the qubit from its environment and can
increase the qubit coherence time. This method is conceptually
similar to the spin-echo method. Specific examples of our
reconstructed Liouvillians including their experimental errors are
presented in figures~\ref{fig2}-\ref{fig4} in comparison to our
theoretical predictions.

\begin{figure}
\includegraphics[width=0.5\linewidth]{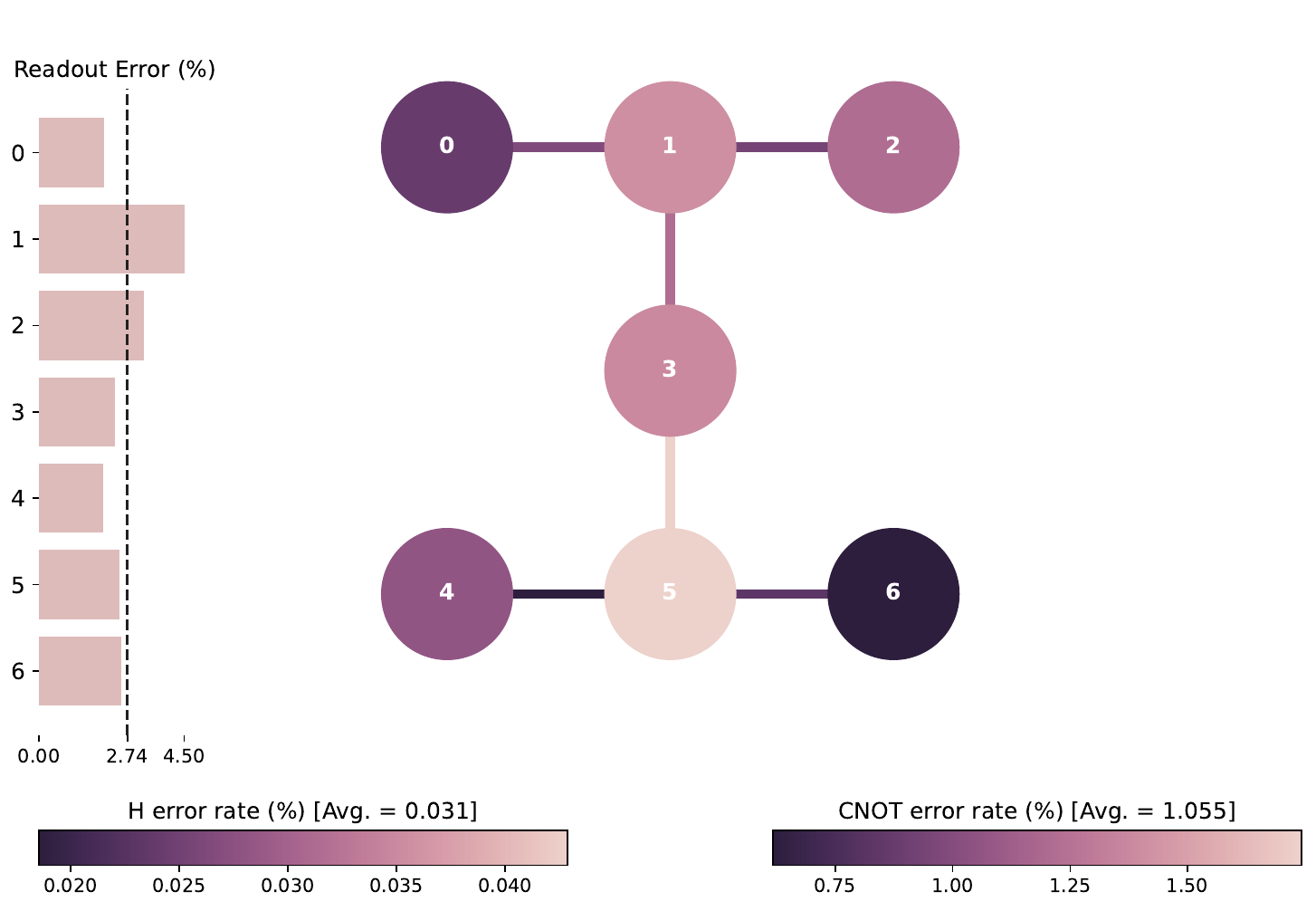}
\includegraphics[width=0.5\linewidth]{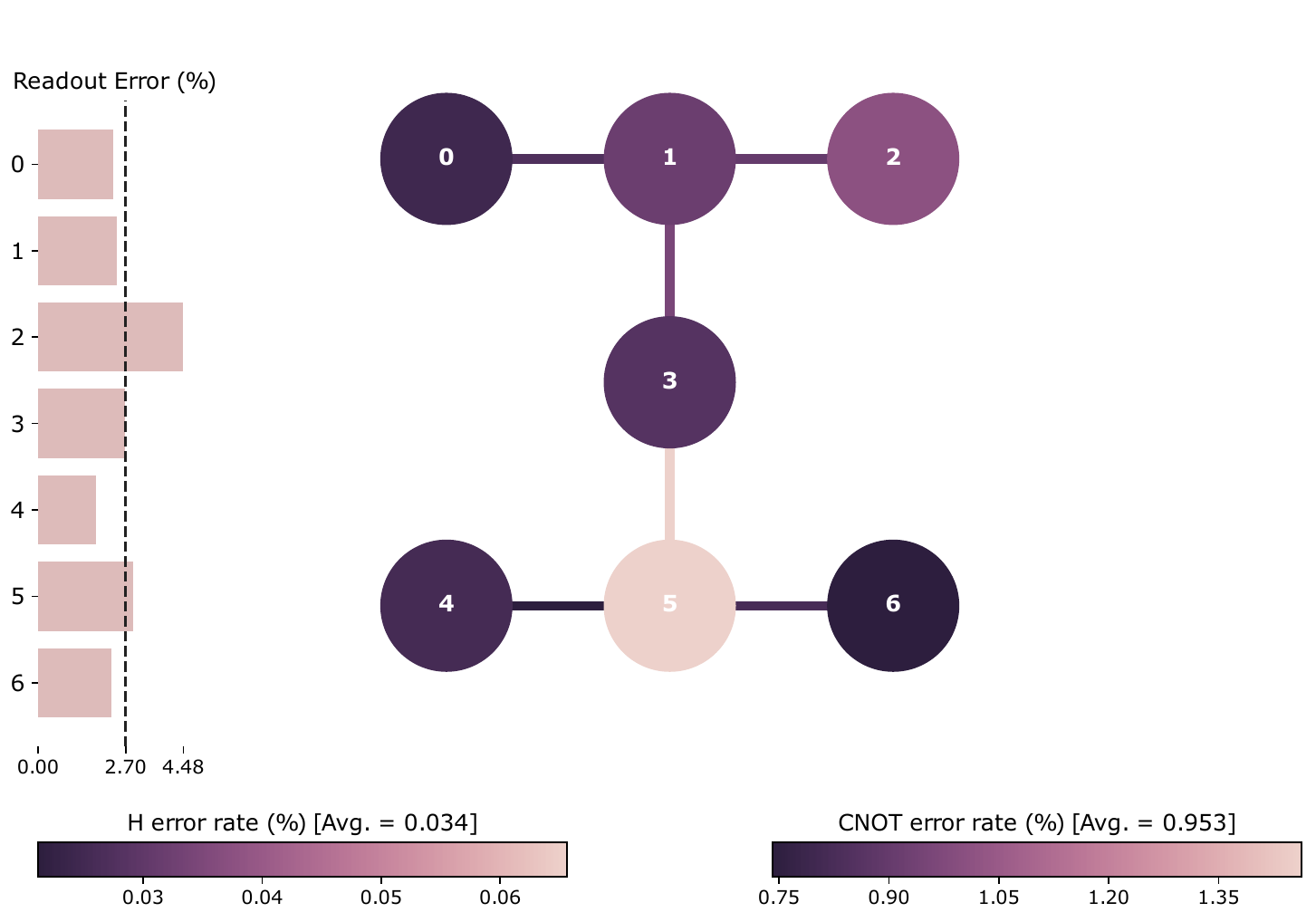}
\caption{Examples of the calibration data of the seven-qubit IBM
quantum processor (i.e., Nairobi) for (a) single- and (b)
two-qubit experiments, as completed on 18 and 22 April, 2023,
respectively. Source:~\cite{IBMQ}. Here, $H$ stands for the
Hadamard gate.} \label{fig5}
\end{figure}

Examples of the coupling map and calibration data for the used
quantum processor are shown in Fig.~\ref{fig5}. Our experiments
were implemented on a seven-qubit IBMQ processor
(Nairobi)~\cite{IBMQ} and we used the Qiskit Runtime environment,
which provides a controllable error mitigation and suppression to
perform our experiment. Note that, we also performed experiments
on several other quantum processors (including the Oslo processor)
from IBM, which we selected based on their coupling maps. However,
the Nairobi processor resulted in the best results compared to our
theoretical predictions and the lowest experimental errors.

\begin{figure}
\includegraphics[width=.48\linewidth]{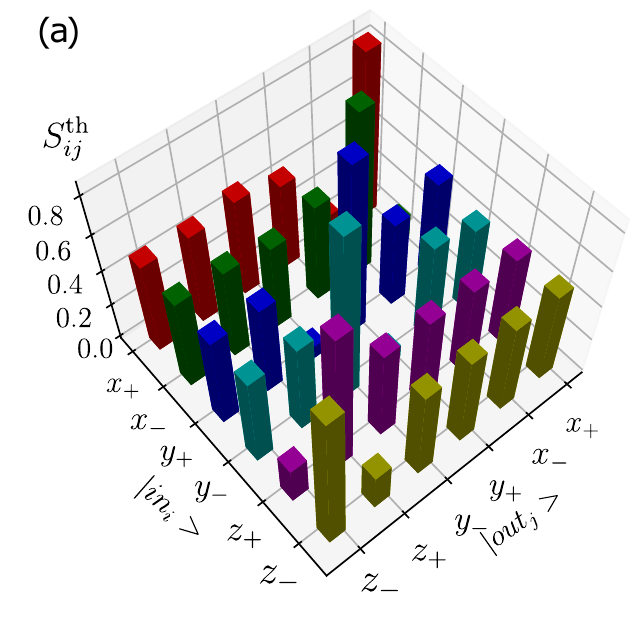}
\includegraphics[width=.48\linewidth]{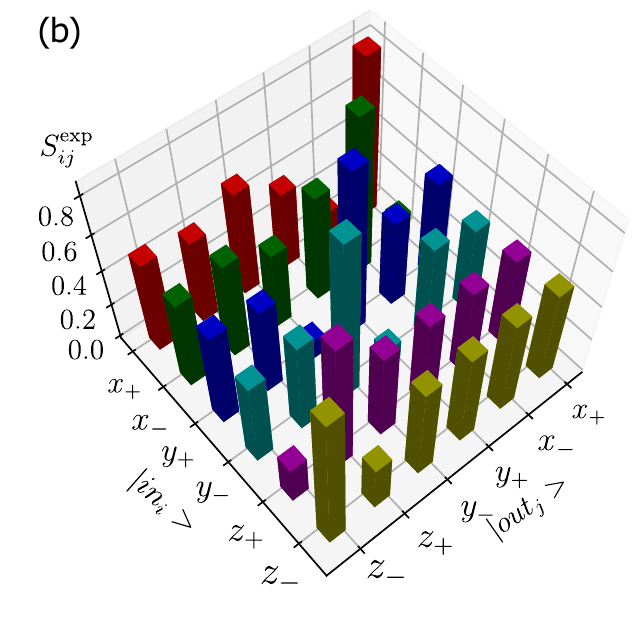}\\

\includegraphics[width=.48\linewidth]{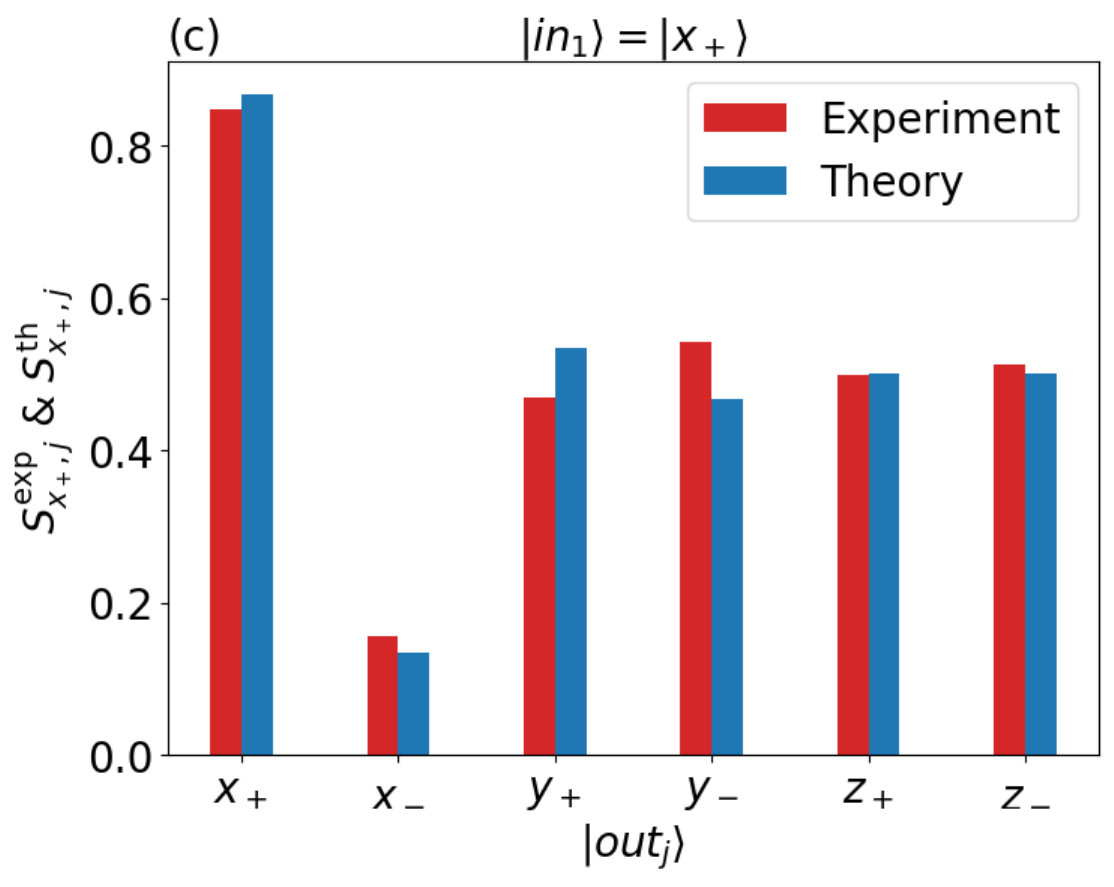}
\includegraphics[width=.48\linewidth]{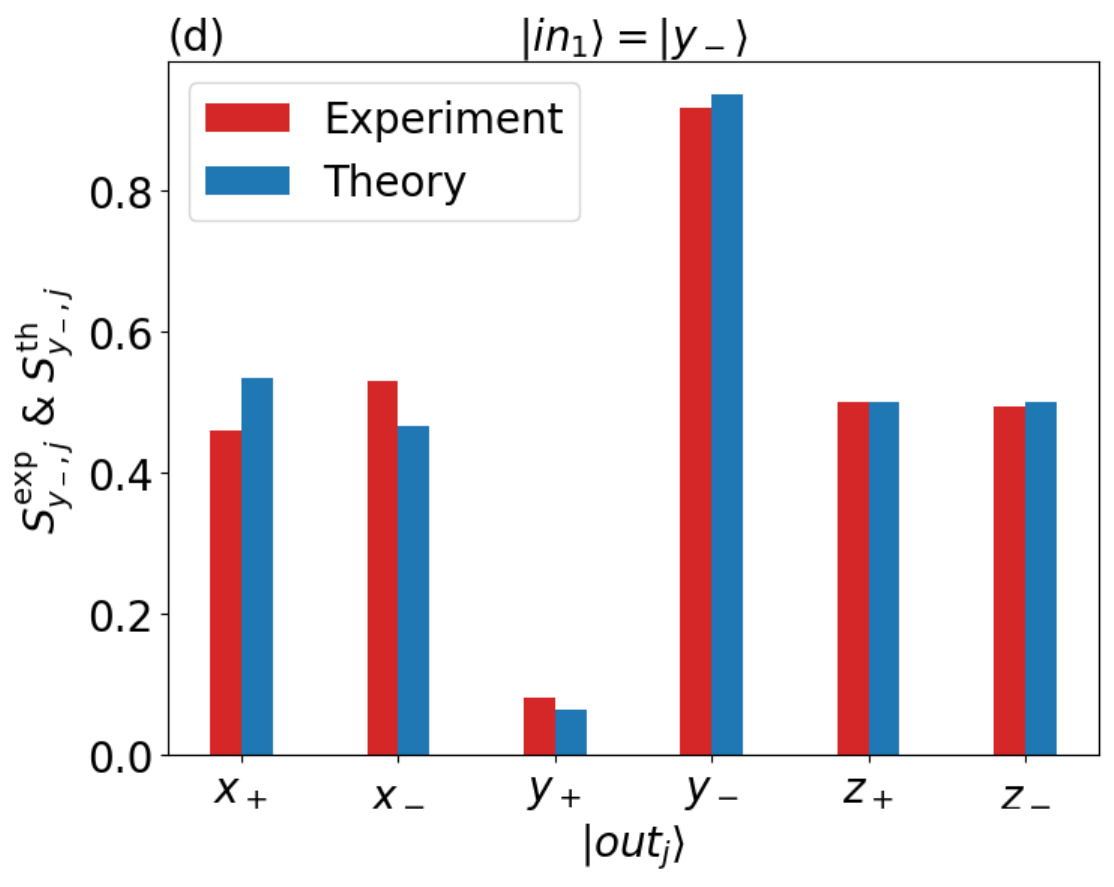}
\caption{Theoretical (a) and experimental (b) matrix elements
$S_{ij}=L_{ij}'dt + 1$, corresponding to the Liouvillian elements
$L_{ij}'$, for our single-qubit experiment performed for $\omega
dt=1/15$ and $\gamma/\omega=0.96$. These matrices correspond to
the first LEP shown in Figs.~\ref{fig4}(a) and \ref{fig4}(b).
Comparison of the cross-sections of the figures in panels (a) and
(b) for chosen input states: $|x_+\rangle$ (c) and $|y_-\rangle$
(d), where the theoretical predictions are represented by blue
bars, and the experimental results by red bars.} \label{fig6}
\end{figure}

\begin{figure}
\includegraphics[width=.48\linewidth]{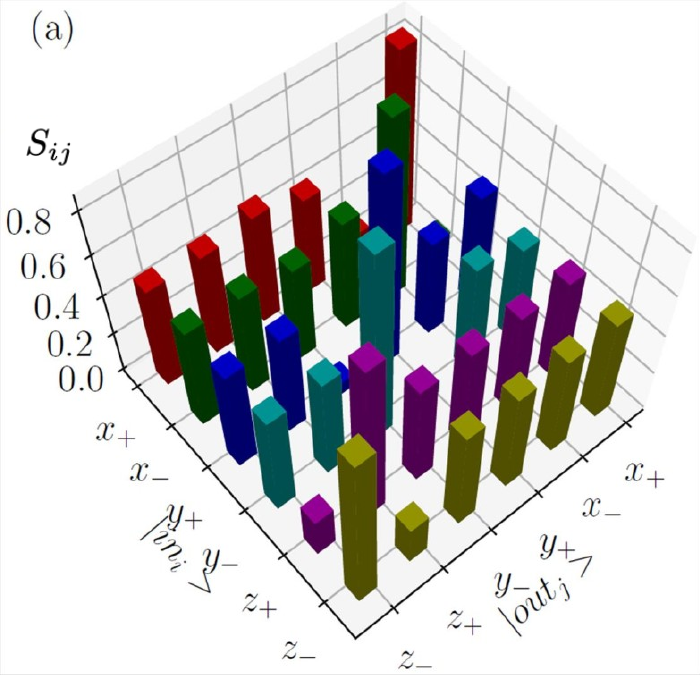}
\includegraphics[width=.48\linewidth]{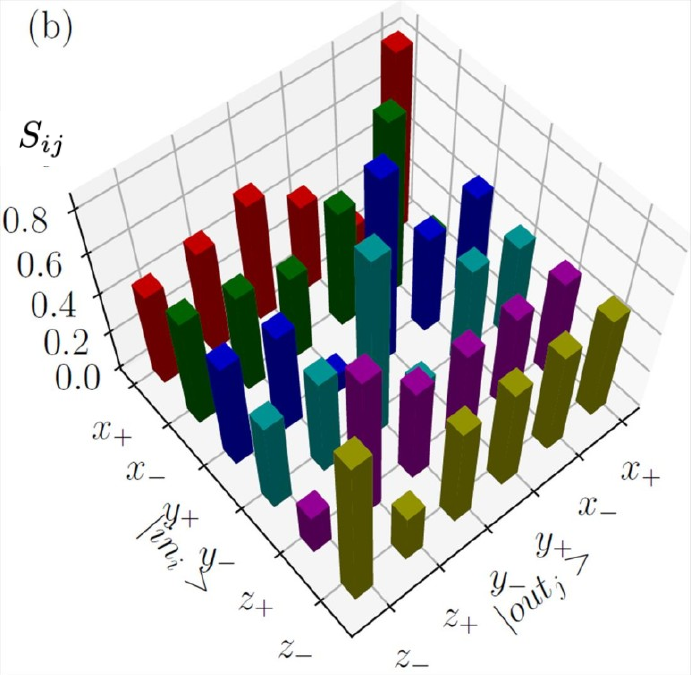}\\

\includegraphics[width=.48\linewidth]{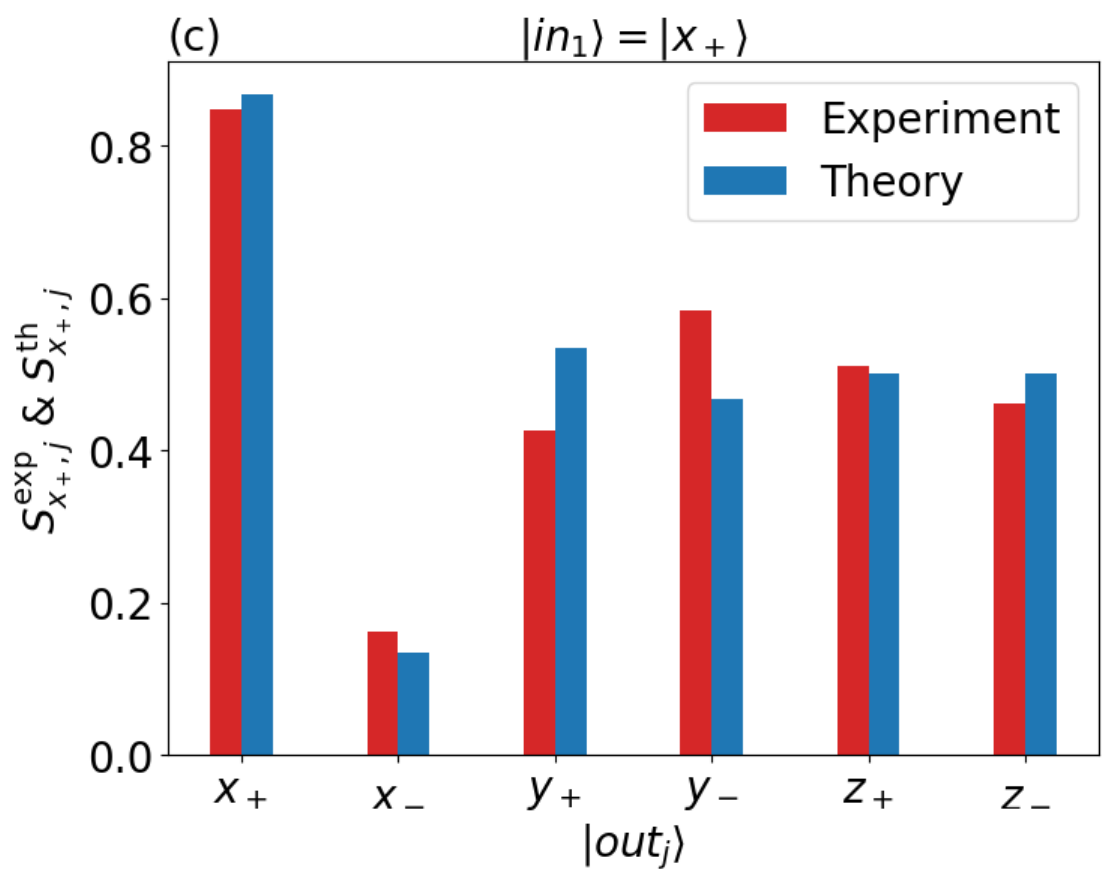}
\includegraphics[width=.48\linewidth]{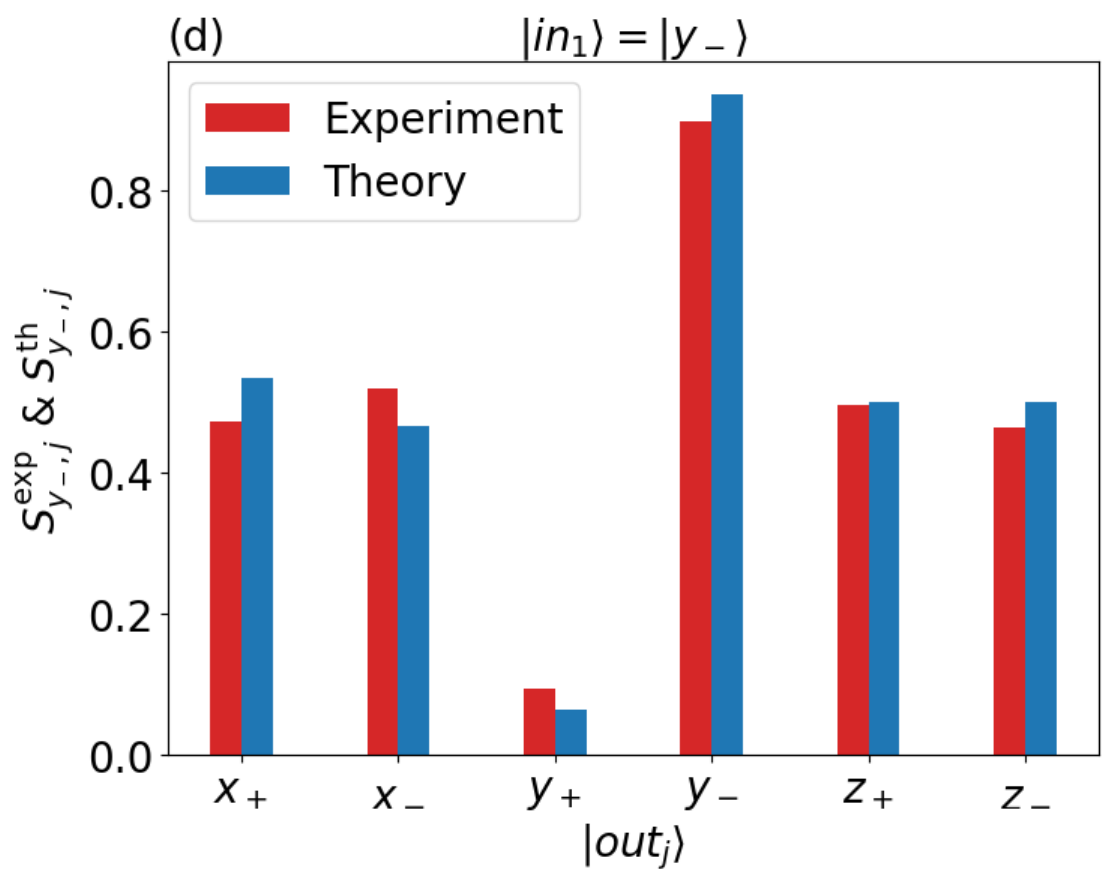}

\caption{Same as in Fig.~\ref{fig6}, but for our two-qubit
experiment. These matrices correspond to the first LEP shown in
Figs.~\ref{fig4}(d) and~\ref{fig4}(e). } \label{fig7}
\end{figure}

Specifically, the theoretical and reconstructed process matrices
$S$, at the first LEP in Figs.~\ref{fig4}(a) and \ref{fig4}(b),
are shown in Fig.~\ref{fig6} obtained for a single-qubit circuit.
Analogously, Fig.~\ref{fig7} shows $S$ obtained for the two-qubit
circuit [depicted in Fig.~\ref{fig1}(b)] for the first LEP in
Figs.~\ref{fig4}(d) and~\ref{fig1}(e). Moreover, in
Figs.~\ref{fig4}(c) and \ref{fig4}(f), we show the scalar products
(overlaps)
$O_{12}^{\exp}=|\langle\tilde\sigma^{\exp}_{1}|\tilde\rho^{\exp}_{2}\rangle|$
between the experimental left and right eigenmatrices,
$\tilde\sigma^{\exp}_{1}$ and $\tilde\rho^{\exp}_{2}$, compared to
$O^{\rm th}_{12} =|\langle\tilde\sigma_{1}^{(0)}
|\tilde\rho_{2}^{(0)}\rangle|$ for the ideal theoretical case. It
is seen that  $\tilde\sigma^{\exp}_{1}$ and
$\tilde\rho^{\exp}_{2}$ are practically coalescent (as
$O_{12}^{\exp}\approx 1$) near $\gamma_x^{\pm}$ confirming the
generation of LEPs. We can conclude that the observed bifurcations
of the experimental eigenvalues at LEPs for $\gamma_x^{\pm}=(2\pm
1)\omega$, and the coalescence of the corresponding eigenmatrices
reconstructed for both the single- and two-qubit experiments are
in good agreement with our theoretical predictions.

In the single-qubit case, we applied the Qiskit optimization
level~1, which enables a single-qubit gate optimization, together
the resilience level~1, which enables readout error mitigation. In
the two-qubit experiments, we used the optimization level~3, which
enables a dynamical decoupling error suppression, together with
the resilience level~1. Moreover, we experimentally tested
different qubits to find that the best results were obtained for:
Qubit \#0 in the single-qubit experiment and Qubits \#4 and \#5
for the two-qubit experiments. See Table~\ref{table3} for more
experimental characteristics of chosen qubits and gates.

In general, our single-qubit experiments are less noisy compared
to the two-qubit and, especially, three-qubit experiments.
However, at a single point, $\gamma_x/\omega=1.379$, we observed a
sudden jump in both the real and imaginary parts of
$\lambda_{1,2}$, with values increasing by three orders of
magnitude compared to other measured values in a single-qubit
experiment. Consequently, this data point is not visible in
Figs.~\ref{fig4}(a) and \ref{fig4}(b) as it lies beyond their axis
range. In contrast, our two-qubit experiment, along with
single-qubit and two-qubit simulations at this point, showed no
such singularity. Therefore, this sudden peak is most likely due
to a malfunction in the experimental setup.

Finally, we note that our experiments for demonstrating the NHH
dynamics are challenging even for simple systems due to highly
entangling three-qubit operations
$\hat{U}^{(3)}=\sum_l\hat{A}_l\otimes\ket{l}\bra{0}$ implemented
by quantum circuits. The results of our three-qubit experiments on
IBMQ processors~\cite{IBMQ} are quite noisy and not shown here.
Although our main experimental results, presented in
Figs.~\ref{fig3} and~\ref{fig4}, are limited to single-, two-, and
three-qubit experiments, they show the potential of QPT for
revealing and manipulating LEPs.
\begin{table}
\begin{tabular}{ |c|c|c|c|c|c|c| }

 \hline
  Qubit   & $\bar T_1 (\mu s)$& $\bar T_2 (\mu s)$& Freq.    & Anharm.       & Single-qubit &  Basic gates\\
          &                   &                   &  (GHz)   & (GHz)         & gate~error   &           \\
 \hline
     \#5  &$109\pm 14$& $76\pm 7$ &  5.18        &-0.34& $2.98\times10^{-4}$ &$I, R_Z, X, \sqrt{X}$, CNOT\\
     \#4  &$155\pm 23$& $20\pm 1$ &  5.29        &-0.34& $2.68 \times 10^{-4}$                &                        \\
 \hline
     \#0  &$126\pm 15$& $33\pm 2$&5.26 &-0.34& $2.27\times10^{-4}$& $I$, $R_Z$, $X$, $\sqrt{X}$ \\
 \hline
\end{tabular}
\caption{Calibration data of the IBMQ Nairobi superconducting
processor used for our experiments with 1 and 2 qubits. Here,
$R_z$ is a single-qubit rotation around the $z$-axis for various
angles, while $\bar T_1$ and $\bar T_2$ are, respectively, the
relaxation and decoherence times of a given qubit averaged for a
few days (i.e., 14--16.04.2023). The CNOT error rate between
qubits \#4 and \#5 is estimated to be $6.5\times10^{-3}$. Qubit
numbers $\#n$ refer to those in Fig.~\ref{fig5}.
Source:~\cite{IBMQ}.} \label{table3}
\end{table}


\section{Physical interpretation of transitions at LEPs}

Tables~\ref{table1} and~\ref{table2} present experimental
eigenvalues depicted in Fig.~\ref{fig4}, which are essential for
interpreting the decay transitions observed at the LEPs. In the
ideal version of our model, LEPs should theoretically occur at the
$\gamma_x/\omega$ values of 1 and 3. However, in our experiments,
these points were shifted. Specifically, for our one-qubit
experiments, the first and second LEPs were observed within the
ranges (0.83,0.97) and (2.90,3.03), respectively. Similarly, in
the two-qubit experiments, the LEPs appeared within the ranges
(0.83,0.97) and (3.31,3.45). As indicated in the tables, achieving
exact tuning to the LEPs was not possible. For instance, in both
experiments, the value $\gamma_x/\omega = 0.97$ already exceeded
the first LEP; this suggests that the LEP is shifted from 1 to a
slightly smaller value on the $\gamma_x / \omega$ axis, by at
least 0.03. Additionally, the second LEP appears shifted from 3 to
a slightly smaller value for the one-qubit experiment, whereas in
the two-qubit experiment, it is shifted to a much higher value by
over 0.31 compared to ideal predictions. Consequently, we were
unable to precisely adjust the system to match the experimental
LEPs; nonetheless, our measurements for both one- and two-qubit
setups are relatively close to these points. For clarity, we refer
to LEPs at approximate $\gamma_x/\omega$ values of 1 and 3,
understanding that these points are slightly shifted in our
experimental configurations.

Before interpreting the observed LEPs and related effects, it is
helpful to revisit both the geometric and physical interpretations
of four types of system evolutions (paths), each associated with a
specific type of a Liouvillian eigenvalue as described in
Refs.~\cite{Baumgartner2008b, Albert2014}: Type 1: For $\lambda_n
= 0$, the corresponding Liouvillian eigenvectors determine the
system's steady state. Type 2: When $\lambda_n < 0$, the system
exhibits exponential decay to zero over time. Type 3: If two
eigenvalues form a complex-conjugate pair, the system follows a
spiraling trajectory as these eigenvalues decay to zero in the
infinite-time limit. These three types of eigenvalues are observed
in our model. For completeness, we also mention Type 4, which
occurs when an eigenvalue is purely imaginary. In this case, the
system exhibits oscillating coherences, following a circular path
geometrically. Here, the eigenmatrix associated with such an
eigenvalue is preserved in the infinite-time limit. However, Type
4 is not observed in our model. Overall, the general evolution of
the system results from the superposition of these distinct types
of evolutions.

As shown in Tables~\ref{table1} and~\ref{table2}, and
Figs.~\ref{fig2}--\ref{fig4}, all four eigenvalues are
non-positive for any value of the damping rate $\gamma_x$.
Specifically, for $\gamma_x/\omega < 1$ and $\gamma_x/\omega > 3$,
the eigenvalues are four distinct real values: one zero and three
negative. In these regions, the system decay at long times is
exponential. In contrast, within the range $\gamma_x/\omega \in
(1,3)$, two of the eigenvalues form a complex-conjugate pair,
$\lambda_1 = \lambda_2^*$, accompanied by one zero and one
negative eigenvalue, as in the previous cases. Here, the decay
rate ${\rm Re}(\lambda_1) = {\rm Re}(\lambda_2)$ is modulated by
an oscillatory (or spiraling) term ${\rm Im}\lambda_1$. When
complex-conjugate pairs with negative real parts are present, the
system is in the ``spiraling regime,'' following a spiral
trajectory (superposed with the exponential decay). Otherwise, it
is in a ``non-spiraling regime,'' where decay is purely
exponential without oscillations.

At the LEPs, where the Liouvillian is not diagonalizable, the
transition between these spiraling and non-spiraling regimes takes
place. The tables also show the Liouvillian spectral gap (also
known as an asymptotic decay rate), commonly defined as the
absolute difference between the real parts of the largest and
second-largest eigenvalues of the Liouvillian. Given that
$\lambda_0 = 0$ here, the spectral gap is simply $|{\rm
Re}(\lambda_1)|$, as given in Tables~1 and 2. In Markovian
processes (as studied here), the spectral gap correlates with the
convergence rate to a steady state, representing the slowest
non-zero rate of convergence to the infinite-time state.
Therefore, a larger spectral gap indicates faster system mixing or
convergence.

Therefore, the experimental observation of the transitions at the
LEPs is not merely a mathematical curiosity; it carries
significant physical implications.

\section{Discussion and conclusions}

We experimentally demonstrated how to engineer and tune a quantum
process to approach and detect quantum LEPs via quantum process
tomography. We believe it is important to highlight here the
following aspects of our experiment in comparison to previous
studies:

1. Numerous experimental studies on HEPs have been documented.
However, to our knowledge, only four experiments, reported in
Refs.~\cite{Chen2021, Chen2022, Zhang2022, Bu2023}, focus on LEPs.
It is important to emphasize that, unlike HEPs, LEPs account for
the effect of quantum jumps~\cite{Minganti2019}. As a result, LEPs
represent true quantum phenomena, whereas HEPs are semi-classical,
although there are quantum systems in which LEPs and HEPs
coincide.

2. For the first time, we experimentally demonstrated a system
exhibiting LEPs without HEPs, thus confirming the prediction made
in Ref.~\cite{Minganti2019} that such systems exist.

3. Furthermore, to our knowledge, we are the first to report
experimental LEPs revealed through QPT. Unlike QST, which
reconstructs only quantum states, QPT allows for the complete
reconstruction of Liouvillians and their associated LEPs. Previous
experimental observations of LEPs in quantum circuits have relied
on QST~\cite{Naghiloo2019, Chen2021, Chen2022}.

We note that QST is used to characterize the output state of a
quantum processor by measuring the output at various measurement
bases assuming that identical states are input to the processor at
each run of the QST protocol. QPT on the other hand is used to
characterize the map or the transformation that the processor
applies to any input state. Here, the processor is probed using
different input states and QST is performed at the output of the
processor for each of these input states. Thus, QST helps
characterize the output state of a processor but does not tell the
inner workings of the processor. QPT on the other hand involves
QST and is used to understand the inner working of the processor.

4. In previous experimental studies of LEPs~\cite{Chen2021,
Chen2022, Zhang2022, Bu2023}, a single qutrit was employed to
simulate the dissipative evolution of a qubit. In contrast, we
use, in particular, two and three qubits on an IBMQ processor to
simulate the evolution of a single qubit, leading to a more
complex system. We then employed tomographic techniques to
reconstruct the experimental Liouvillians and their LEPs using
both single- and two-qubit operations.

5. We acknowledge that there are experimental studies on QPT for
single superconducting qubits (as discussed below). However, these
studies did not reveal any LEPs, as LEPs can only be observed with
specifically designed damping and/or amplification channels using
precisely tuned parameters. In contrast, our experiment utilized
additional qubits to implement the desired damping channels,
enabling the observation of LEPs.

Let us emphasize again that QPT is a well-established method
frequently applied to systems like linear-optical setups and
trapped ions. However, its applications in circuit QED systems
have largely focused on characterizing specific aspects of device
performance, such as testing quantum gate fidelities (see, e.g.,
Refs.~\cite{Bialczak2010, Rodionov2014, Shukla2020}).
Consequently, QPT has often been replaced by other quantum
characterization techniques, including QST, gate set tomography,
or randomized benchmarking (see, e.g., Ref.~\cite{Eisert2020} and
references therein). Only in a few recent experiments with
superconducting qubits, QPT (in generalized forms) has been
employed to fully characterize dissipative
dynamics~\cite{Samach2022, Pears2023}. Specifically, in
Ref.~\cite{Samach2022}, a refined QPT technique known as Lindblad
tomography was developed and used to reconstruct an idling
channel, fully characterizing the natural (non-engineered) noise
of a superconducting quantum processor. In contrast, our work
applies QPT to fully characterize the dynamics of qubits with
\emph{specifically engineered} dissipation channels. More
importantly, to our knowledge, QPT has not been applied at EPs
before. Various studies (see, e.g., Refs.~\cite{Langbein2018,
Mortensen2018, Lau2018, Wolff2019, Zhang2019, Chen2019,
Naikoo2023, Loughlin2024} and references therein) showed that
noise at EPs can significantly increase, making it unclear, prior
to our experiment, whether reliable results could be obtained
through QPT at these quantum singularities not only on circuit-QED
platform, but on any experimental platform. We have successfully
demonstrated that quantum LEPs can indeed be reliably and fully
revealed experimentally via QPT.

While QPT enables complete identification and characterization of
LEPs, it is not the only approach. QPT can indeed be substituted
by alternative approaches such as: (i) topological engineering
through encircling an LEP, (ii) Lindblad tomography, or (iii)
Heisenberg-Langevin methods. (i) Topological techniques are widely
applied to reveal and characterize HEPs in classical physics (see,
e.g., Refs.~\cite{Bergholtz2021, Ashida2020, Ergoktas2022} and
references therein). Experimental observations of an LEP, which
can be considered a form of topological engineering, have been
recently reported in Ref.~\cite{Bu2023}, where the enhancement of
a quantum Otto heat engine through encircling an LEP was observed.
In contrast, (ii) Lindblad tomography, which is a modified QPT
method introduced recently in~\cite{Samach2022}, has yet to be
employed for LEP identification. Finally, (iii) LEPs can also be
fully characterized using the Heisenberg-Langevin equations of
motion, as studied theoretically in~\cite{Perina2022, Perina2023,
Wakefield2024, Perina2024, Thapliyal2024}, though these approaches
for detecting LEPs in dissipative systems remain experimentally
untested.

It is also notable that earlier LEP experimental demonstrations
relied not only on reported experiments but on preliminary trials
that located these singularities and validated theoretical
predictions. In our approach, we reveal LEPs solely using the
physical system, independent of a pre-existing theoretical model.
This process is illustrated in Figs.~\ref{fig3}(b), where we first
obtained experimental data and subsequently developed and applied
a theoretical simulation shown in Fig.~\ref{fig2}(b).

LEPs are distinguished by both degenerate Liouvillian eigenvalues
and coalescent eigenmatrices, underscoring the usefulness of
Liouvillian tomography in distinguishing LEPs from Liouvillian
diabolical points. Although an LEP can also be identified by
measuring topological properties around it \cite{Bu2023} (See also
topological methods developed for detecting HEPs, such as those
discussed in Ref.~\cite{Egenlauf2024} and references therein),
partial QPT is still highly beneficial. It aids in pinpointing a
potential LEP, guiding optimal paths to encircle and reveal its
topological properties.

Regarding our experiments, the operations for all the experiments
are automatically transpiled into a sequence of single- and
two-qubit gates, which were physically implemented on a given
quantum processor. The physical qubits were selected based on
their quality and connectivity, which contribute to the optimal
performance of the quantum circuit. To suppress the noise even
more, we have explored the state-of-the-art noise-canceling
techniques for quantum processors. While we applied experimentally
various methods, we found the dynamical decoupling technique to be
the most useful. Various QPT methods, which are equivalent under
ideal measurement conditions, can be used for revealing LEPs as
discussed in Appendix~C. However, we observed that the least
perturbed Liouvillians for our experimental data were obtained for
the QPT method described above (Method~1).

Our experiment and data postprocessing were specifically designed
to reconstruct Liouvillian (non-Hermitian) dynamics, not those of
an NHH. However, the Hermitian Hamiltonian $\hat{H}$ and the
quantum jump operators $\{\hat{L}_{\mu}\}$, and thus also the
corresponding NHH for the studied system could, in principle, be
determined using the above-mentioned Lindblad tomography, as
demonstrated in a circuit-QED system in Ref.~\cite{Samach2022}.
Lindblad tomography is an adaptation of Liouvillian tomography
(i.e., standard QPT), specifically aimed at separately
reconstructing the Lindbladian dissipators and Hamiltonian of a
given process, rather than directly reconstructing its full
Liouvillian. This method relies on a series of pre- and
post-pulses to perform rotations needed to reconstruct a channel
(or process) at each discrete time step, consistent with standard
QPT; and practically the same approach we employed in our
experiments. The experimental data, encompassing all combinations
of pre- and post-pulses and channel durations, are then processed
using a classical optimizer based on maximum likelihood
estimation, as detailed in \cite{Samach2022}. This postprocessing
step in Lindblad tomography differs from standard QPT (i.e.
Liouvillian tomography), including the approach we followed, in
that we did not apply this type of postprocessing for NHH
reconstruction. This is because our primary focus was on
demonstrating LEPs, rather than experimentally verifying the
absence of HEPs, which we assumed. A strong agreement between the
theoretical and experimentally reconstructed Liouvillians,
particularly for the single- and two-qubit circuits, also provides
indirect validation of our assumption regarding the form of the
corresponding NHH and the absence of its HEPs.

We implemented our model using unitary gates, as only unitary
operations can be executed on IBMQ. However, the model can also be
realized on different platforms, particularly in a linear-optical
system. The advantage of such systems is that damping channels can
be directly applied to a single qubit, without the need for
auxiliary qubits to simulate dissipative evolution. Single- and
two-qubit QPT methods can be readily applied to linear-optical
systems, where qubits are encoded in photon polarization.
Moreover, controllable damping channels can be implemented through
various techniques (see, e.g., Refs.~\cite{Almeida2007, Ku2022}
and references therein), making it feasible to realize the current
or alternative models exhibiting LEPs in linear-optical systems.
This approach could also enable a more precise reconstruction of
the Liouvillian, due to significantly lower noise levels of
optical systems compared to circuit QED systems.

Regarding experimental feasibility beyond two or three qubits, we
emphasize that the IBMQ processors used in our experiments
(specifically Nairobi and Oslo) were relatively noisy, which
hindered our ability to obtain clear evidence of LEPs with the
three-qubit circuits for the studied model. The fidelity achieved
was much lower than that of single- and two-qubit circuits, as
illustrated by comparing Fig.~\ref{fig3}(a) with
Figs.~\ref{fig4}(a) and~\ref{fig4}(d). In contrast, quantum
computing platforms based on trapped ions, such as those from
IonQ, Honeywell, or Alpine Quantum Technologies (AQT), generally
exhibit significantly lower noise levels. Therefore, implementing
QPT to reveal and characterize LEPs in larger systems with more
qubits appears more feasible on such a platform.

For potential further applications of QPT in identifying LEPs or
their generalizations, it is noteworthy that Lin et al.
\cite{Lin2024} has recently generalized the concept of quantum EPs
from Markovian to non-Markovian dynamics. Their approach utilizes
exact methods for non-Markovian dynamics, specifically the
pseudomode equation of motion and hierarchical equations of motion
(see, e.g., Ref.~\cite{Lambert2019} and references therein). Both
methods rely on auxiliary degrees of freedom, which can be
implemented through additional qubits or by utilizing
higher-energy levels in superconducting systems, such as
transmons. These auxiliary systems effectively increase the
dimensionality of the superoperators being studied, while
fundamentally allowing for the application of QPT to identify
quantum EPs in non-Markovian systems. The pseudomode approach is
particularly analogous to the standard Markovian master equation,
possessing a Lindblad-type structure. Thus, as proposed in
\cite{Lin2024}, by defining an extended Liouvillian superoperator,
non-Markovian EPs can be characterized by the degeneracies in its
complex spectrum. This framework appears well-suited for
experimentally generating EPs in non-Markovian systems and
revealing them via QPT. Notably, there is no fundamental
distinction between applying QPT to the standard master equation
and the generalized version based on pseudomodes. The key
difference lies in the use of additional auxiliary qubits and/or
qudits to simulate the desired non-Markovian dynamics. While there
appears to be no fundamental obstacle to applying a modified QPT
for revealing LEPs in non-Markovian systems, at least within a
few-qubit framework, a more in-depth analysis is necessary to
develop the specific details for implementing QPT of
experimentally simulated dynamics using IBMQ systems. Moreover,
the challenge arises with the increased noise associated with a
larger number of qubits. Consequently, implementing such
non-Markovian EPs on IBMQ presents significant difficulties as we
encounter in the present study. In contrast, utilizing a
trapped-ion quantum computer (as mentioned above) could provide a
more favorable platform for realizing these dynamics, given their
lower experimental noise.

To highlight potential future studies on LEPs and underscore the
importance of this work compared to HEPs, it is worth noting that
HEPs have already garnered significant interest for their unique
topological and other distinctive properties. This research into
HEPs holds value not only for fundamental reasons, such as
studying novel types of quantum phase transitions, but also for
potential advanced applications, including quantum sensing.
However, HEPs rely on a semi-classical dissipation model that
omits quantum jumps, potentially leading to fundamental issues,
such as the preservation of canonical commutation relations. For
small quantum systems, a correct dissipation model (at least
without postselection on quantum trajectories) must account for
quantum jumps. Consequently, LEPs, as natural extensions of HEPs
with proper inclusion of quantum jumps, should be used in place of
HEPs. Initial theoretical and experimental results of other groups
suggest that encircling an LEP could enhance the efficiency of
quantum heat engines~\cite{Zhang2022, Bu2023}. While further
research is essential to establish their full potential for
improving quantum sensing and the performance of quantum heat
engines when encircling LEPs in small quantum systems, we believe
such investigations are highly valuable. With future experimental
investigations into LEP applications in mind, we have demonstrated
that QPT is a practical and effective tool for characterizing,
verifying, and validating the dynamics of few-qubit systems near
LEPs through precise engineering and control of dissipation
channels.

In discussing potential future work, it is worth mentioning
Liouvillian diabolical points (LDPs). While LDPs are not the focus
of the current work, they can be uniquely revealed by QPT and
their occurrence signals intriguing physical phenomena, such as
Liouvillian spectral collapse \cite{Minganti2021a} predicted
within the Scully-Lamb laser model. Note that HEPs and LEPs for
systems described by the Scully-Lamb model have also been
predicted \cite{Arkhipov2019, Arkhipov2020a}, and HEPs have even
been observed \cite{Peng2014b, Chang2014}.

In summary, since their introduction five years ago, LEPs have
garnered increasing interest. The majority of LEP studies remain
theoretical, with only five experimental demonstrations to date of
LEPs (and only in single qutrits) \cite{Naghiloo2019, Chen2021,
Chen2022, Zhang2022, Bu2023}, underscoring the importance of
experimental investigations in this field. We believe our work not
only presents the first observations of LEPs in one-, two-, and
three-qubit systems but also demonstrates the feasibility of
process tomography as a universal method for detecting LEPs.


\appendix

\section{Matrix representation of superoperators}

To understand the basic idea of LEPs and their relation to QPT, we
recall the matrix formalism of superoperators which applies to
Liouvillians. A general matrix $\hat{O}$ can be formally
vectorized (or flattened) with a function ${\cal F}$ as
    \begin{equation}
    \hat{O}=\sum_{m,n} O_{mn} \ket{m}\bra{n}\rightarrow
     \ket{\tilde O}=\Vec{\hat{O}} = \sum_{m,n} O_{mn} \ket{m}\otimes \ket{n^*},
    \label{flatten1}
    \end{equation}
where $*$ denotes complex conjugate and, for clarity, flattened
quantities are henceforth marked by tilde. Thus, a single-qubit
matrix $\hat{\rho}$ can be flattened as
\begin{equation}
    \hat{\rho}=
    \left[\begin{array}{rr}
    \rho_{00}& \rho_{01} \\
    \rho_{10}& \rho_{11}
    \end{array} \right]
    \longrightarrow \ket{\tilde\rho}=\Vec{\hat{\rho}} =
         [\rho_{00},\rho_{10}, \rho_{01},\rho_{11}]^T,
     \label{flatten2}
    \end{equation}
where $T$ denotes transposition. The inverse function
$\Invec{\ket{\tilde\rho}}$ gives the standard form of the density
matrix $\hat{\rho}$. Arbitrary right-hand-side (RHS) and
left-hand-side (LHS) acting superoperators, say ${R}[\hat{O}_1]$
and ${L}[\hat{O}_1]$, can be represented by matrices
$\tilde{R}[\hat{O}_1]$ and $\tilde{L}[\hat{O}_1]$, defined,
respectively, as:
    \begin{eqnarray}
    \tilde{R}[\hat{O}_1] \ket{\tilde O}_{2} &=& (\mathds{1}_N\otimes \hat{O}_1^{T})\ket{\tilde O}_{2}, \nonumber \\
    \tilde{L}[\hat{O}_1] \ket{\tilde O}_{2} &=& (\hat{O}_1\otimes \mathds{1}_N)\ket{\tilde O}_{2},
    \label{Eq:RightleftAction}
    \end{eqnarray}
 where
$\mathds{1}_N$ is the identity operator of dimension $N={\rm
size}(\hat{O}_1)$. By applying this convention, the Liouvillian in
the Linblad master equation can be represented as
\begin{eqnarray}
    \tilde{\LL}&=&  -i
    \left(\hat{H}\otimes \mathds{1}_N  - \mathds{1}_N\otimes \hat{H}^{T}\right)
\\
    &&+\sum_n  \hat{\Gamma}_{n}\otimes
    \hat{\Gamma}_n^*
-\frac{1}{2}\left(\hat{\Gamma}_n^\dagger \hat{\Gamma}_n\otimes
        \mathds{1}_N -  \mathds{1}_N \otimes  \hat{\Gamma}_n^{T}
        \hat{\Gamma}_n^*\right),\nonumber
     \label{Lmatrix1}
    \end{eqnarray}
or, equivalently,
    \begin{eqnarray}
    \tilde{\LL}&=&  -i
    \left(\hat{H}_{\rm eff}\otimes \mathds{1}_N
     - \mathds{1}_N\otimes \hat{H}_{\rm eff}^{T}\right)
    +\sum_n  \hat{\Gamma}_{n}\otimes
    \hat{\Gamma}_n^*,
     \label{Lmatrix2}
    \end{eqnarray}
in terms of the effective Hamiltonian $\hat{H}_{\rm eff}$ defined
in Eq.~(\ref{H_eff}). The last term in Eq.~(\ref{Lmatrix2})
represents the effect of quantum jumps on the system evolution.
And this effect can be decreased or even completely removed by a
proper postselection of quantum trajectories, as described by a
hybrid Liouvillian formalism~\cite{Minganti2020}.

\section{Equivalent QPT methods for finding exceptional points}

LEPs can be calculated via the standard superoperator formalism as
described in Ref.~\cite{Minganti2019}. Here we consider three
methods of finding LEPs via QPT for a single qubit. The methods
are equivalent assuming ideal measurements.

\subsection*{Method 1}

From an experimental point of view, it is convenient to determine
LEPs via the QPT based on $6\times 6$ projectors, i.e., assuming
that the input and output states (projections) are the eigenstates
of all the Pauli operators ($i,j=x_+,x_-,y_+,y_-,z_+,z_-$):
\begin{equation}
  |\In_i\rangle,|\Out_j\rangle \in
  \{|x_+\rangle,|x_-\rangle,|y_+\rangle,|y_-\rangle,|z_+\rangle,|z_-\rangle\},
  \label{PauliEigenstates}
\end{equation}
where
$|x_{\pm}\rangle = \frac{1}{\sqrt{2}}\left(|0\rangle \pm |1\rangle\right),$
 $ |y_{\pm}\rangle = \frac{1}{\sqrt{2}}\left(|0\rangle \mp i|1\rangle\right)$, $|z_+\rangle\equiv|0\rangle,$ and $|z_-\rangle\equiv|1\rangle.$
These are arguably the most popular projectors used for QST and
QPT of photon polarization qubits, but can also be applied to
transmon qubits. Thus, for an amplified-dissipative process,
described by the Lindblad master equation with a given Liouvillian
$\LL$, one can measure all its elements
\begin{equation}
   L'_{ij}=\langle \Out_j|\LL\left(\hat\rho=|\In_i\rangle\langle \In_i|\right)|\Out_j\rangle,
  \label{Lij2}
\end{equation}
and, thus, we can reconstruct the $6\times 6$ transformation
matrix $L'$, which represents $\LL$.

\subsection*{Method 2}

LEPs can also be calculated via the QPT for all the Pauli
operators ($k=x,y,z$), i.e.,
\begin{eqnarray}
 \hat\sigma_k&=&|k_+\rangle\langle k_+|-|k_-\rangle\langle k_-|,
 \\
 \hat\sigma_0&=&|z_+\rangle\langle z_+|+|z_-\rangle\langle z_-|=\openone,
\label{eq:Pauli}
\end{eqnarray}
which can be obtained via the projections on their eigenstates,
given in Eq.~(\ref{PauliEigenstates}). Thus, by measuring all the
elements ($m,n=0,...,3$):
\begin{equation}
  L''_{mn}=\frac{1}{2} \Tr\left[\LL\left(\hat\sigma_m\right)\hat\sigma_n\right],
  \label{N6}
\end{equation}
where $\hat\sigma_1\equiv\hat\sigma_x,$
$\hat\sigma_y\equiv\hat\sigma_2,$ and
$\hat\sigma_3\equiv\hat\sigma_z,$ we can reconstruct the $4\times
4$ Liouvillian matrix $L''$ representing $\LL$.

\subsection*{Method 3}

Formally the simplest approach to find LEPs is via the QPT based
on the following $4\times 4$ non-Hermitian input/output projectors
($k,l=1,...,4$):
\begin{equation}
  \hat\rho_{\In, k},\hat\rho_{\Out, l} \in \{|0\rangle\langle 0|,|0\rangle\langle 1|,
  |1\rangle\langle0|,|1\rangle\langle1|\}.
  \label{N3}
\end{equation}
By measuring all the elements
\begin{equation}
  L_{kl}=\Tr\left[\LL\left(\hat\rho_{\In, k}\right)^\dagger \hat\rho_{\Out, l}\right],
\label{eq:L}
\end{equation}
one can reconstruct the $4\times 4$ transformation matrix $L$,
which represents $\LL$. The method, although formally
straightforward, is usually experimentally challenging, and it is
not applied here.

\section{Equivalence of Methods 1, 2, and 3 under ideal
experimental conditions}

Here we briefly demonstrate that all the Liouvillian matrices $L$,
$L'$, and $ L''$ have the same eigenspectra (up to trivial zero
values). It means that the exceptional points corresponding to the
nonzero eigenvalues of $L$, $L'$, and $L''$ are of the same order.
More details on this equivalence are given in Supplement~1.

The spectral decomposition of $L$ reads $L=UAV,$ where $A$ is a
diagonal matrix of the singular values of $L$ and the matrices $U$
and $V$ are unitary matrices constructed from the left and right
eigenmatrices. It can be verified via direct calculations and
using the linearity of the Liouvillians $L$ and $L''$ that $L =
U'' L'' (U'')^\dagger,$ where the respective unitary matrix reads
\begin{eqnarray}
U'' =\frac{1}{\sqrt{2}}
 \left[
\begin{array}{rrrr}
   1 & 0 &  0 &  1 \\
   0 & 1 & -i &  0 \\
   0 & 1 &  i &  0 \\
   1 & 0 &  0 & -1
\end{array}
 \right].
\end{eqnarray}
Hence,
\begin{equation}
  L'' = [(U'')^\dagger U] A [V U''] = (U'')^\dagger L U''
  \label{Ldecomp}
\end{equation}
is a spectral decomposition of $L''$ with the same eigenspectrum
as $L$ given by $A$, but with in general different left and right
eigenmatrices given by unitary matrices $(U'')^\dagger U$ and
$V(V'')^\dagger$. Similarly, we can write the transformation
\begin{equation}
  L''= U' L' (U')^T,
  \label{L2L1}
\end{equation}
where
\begin{eqnarray}
U' =\frac{1}{\sqrt{2}}
 \left[
\begin{array}{rrrrrr}
   1 & 1 &  0 &   0 &  0 &   0\\
   0 & 0 &  1 &  -1 &  0 &   0\\
   0 & 0 &  0 &   0 &  1 &  -1\\
   1 & -1 &  0 &   0 &  0 &   0
\end{array}
 \right].
\end{eqnarray}
We note that $U^T$ and $(U')^T$ are the respective pseudoinverse
matrices to $U$ and $U'$, respectively. In Supplement~1, by
deriving and applying the unitary version of $U'$, we demonstrate
that $L''$ and $L'$ have equal ranks. We also demonstrate that
$L''$ and $L'$ have the same nontrivial eigenvalues. Thus, up to
the two trivial eigenvalues of $L'$, the spectra of $L$, $L'$, and
$L''$ coincide. This is confirmed by our numerical calculations.

Thus, we have demonstrated that the three methods are formally
equivalent, assuming perfect measurements. However, it is
important to stress that under realistic measurement conditions,
the methods usually lead to slightly different reconstructions of
Liouvillians and their LEPs. Analogously, different QST methods,
which are formally equivalent under perfect measurement
conditions, become inequivalent in realistic situations.
Consequently, this leads to varying reconstructions of
experimental density matrices (for comparative studies of QST
methods based on their condition numbers see
Refs.~\cite{Miranowicz2014tomo, Bartkiewicz2016tomo,
Miranowicz2015tomo, Kopciuch2024}).

\section{Estimation of errors in the eigenspectra of
experimental Liouvillians} \label{PerturbedL}

Here, we estimate, based on the derivation of
Ref.~\cite{James2001}, how much experimental (or numerical)
perturbations can affect the eigenspectra of an Liouvillian
superoperator. Let us consider an experimental Liouvillian
${L}^{\exp}$, with the eigenspectra,
\begin{eqnarray}
{L}^{\exp}|\tilde\rho_{n}^{\exp}\rangle &=&
\lambda^{\exp}_{n}|\tilde\rho_{n}^{\exp}\rangle ,\nonumber \\
\langle\tilde\sigma_{n}^{\exp}|{L}^{\exp} &=& \lambda^{\exp}_{n}
\langle\tilde\sigma_{n}^{\exp}|, \label{eigLexp}
\end{eqnarray}
which slightly differ from the spectra of an ideal Liouvillian
${L}_{0}$,
\begin{eqnarray}
{L}_{0}|\tilde\rho_{n}^{(0)}\rangle &=&
\lambda^{(0)}_{n}|\tilde\rho_{n}^{(0)}\rangle ,\nonumber \\
\langle\tilde\sigma_{n}^{(0)}|{L}_{0} &=& \lambda^{(0)}_{n}
\langle\tilde\sigma_{n}^{(0)}|. \label{eigL0}
\end{eqnarray}
So, one can write
\begin{equation}
{L}^{\exp} = {L}_{0}+\delta {L}, \label{Lexp2}
\end{equation}
assuming that $\delta {L}$ is a small perturbation. Note that the
completeness relation,
$\sum_{n}|\tilde\rho^{(0)}_{n}\rangle\langle\tilde\sigma^{(0)}_{n}|
= \mathds{1}_4$, and the orthonormality condition,
$\langle\tilde\sigma_{n}^{(0)}|\tilde\rho_{m}^{(0)}\rangle =
\delta_{nm}$, are satisfied for ${L}_{0}$, and analogously for the
corresponding eigenmatrices of ${L}^{\exp}$. Thus, if  the
Liouvillians are diagonalizable (i.e., apart from their LEPs), we
have
\begin{eqnarray}
f({L}_{0}) &=& \sum_{i} f(\lambda^{(0)}_{i})
|\tilde\rho^{(0)}_{i}\rangle\langle\tilde\sigma^{(0)}_{i}|,
\label{L_expansion1} \\
f({L}^{\exp}) &=& \sum_{i} f(\lambda^{\exp}_{i})
|\tilde\rho^{\exp}_{i}\rangle\langle\tilde\sigma^{\exp}_{i}|,
\label{L_expansion2}
\end{eqnarray}
for any well-behaved functions $f$ of the Liouvillians. Assuming
small perturbations in ${L}$ and in related quantities, we
consider their power-series expansions:
\begin{eqnarray}
{L}^{\exp} &=& {L}_{0}+ \epsilon L_1+\ldots,
\nonumber\\
\lambda^{\exp}_{n} & = & \lambda^{(0)}_{n} + \epsilon\lambda^{(1)}_{n}+\ldots, \nonumber\\
|\tilde\rho^{\exp}_{n}\rangle & = & |\tilde\rho^{(0)}_{n}\rangle +
\epsilon |\tilde\rho^{(1)}_{n}\rangle+\ldots,\nonumber\\
\langle\tilde\sigma^{\exp}_{n}| & = &
\langle\tilde\sigma^{(0)}_{n}| + \epsilon
\langle\tilde\sigma^{(1)}_{n}|+\ldots, \label{expansions}
\end{eqnarray}
in some perturbation parameter $\epsilon$. For simplicity,
hereafter, we omit all the terms with higher powers of $\epsilon$.
So, we can assume that $\delta L\approx \epsilon L_1$. By
inserting these expansions into Eq.~(\ref{eigLexp}), one instantly
obtains Eq.~(\ref{eigL0}) for all the terms independent of
$\epsilon$. Moreover, by collecting all the terms proportional to
$\epsilon$ in these equations, one obtains:
\begin{eqnarray}
F_n^{(0)}|\tilde \rho^{(1)}_{n}\rangle & =& -F_n^{(1)}|\tilde
\rho^{(0)}_{n}\rangle,
 \label{aux1}\\
 \langle\tilde \sigma^{(1)}_{n}|F_n^{(0)}
 &=&-F_n^{(1)}\langle\tilde \sigma^{(0)}_{n}|,
 \label{aux2}
\end{eqnarray}
where $F_n^{(k)}={L}_{k}-\lambda^{(k)}_{n}{\openone_4}$ for
$k=0,1$. Multiplying Eq.~(\ref{aux1}) by
$\langle\tilde\sigma_{n}|$ from the LHS, one obtains
$\lambda^{(1)}_{n} = \langle\tilde\sigma_{n}^{(0)}| L_1
|\tilde\rho_{n}^{(0)}\rangle$, or, equivalently,
\begin{eqnarray}
\delta \lambda &\equiv& \lambda^{\exp}_{n} - \lambda^{(0)}_{n}
\approx \epsilon \lambda^{(1)}_{n} =
\langle\tilde\sigma_{n}^{(0)}|( \epsilon L_1
)|\tilde\rho_{n}^{(0)}\rangle \nonumber \\
&\approx&\langle\tilde\sigma_{n}^{(0)}| \delta{L}
|\tilde\rho_{n}^{(0)}\rangle.
 \label{delta_lambda2}
\end{eqnarray}
By applying Eq.~(\ref{L_expansion1}), with $f({L}_{0})=F_n^{(0)}$,
to Eq.~(\ref{aux1}), one obtains
\begin{eqnarray}
|\tilde\rho^{(1)}_{n}\rangle &\approx& -\sum_{i\;(i\neq n)}
\left(\frac{ \langle\tilde\sigma^{(0)}_{i}| {L}_1
|\tilde\rho^{(0)}_{n}\rangle}{\lambda^{(0)}_{i}-\lambda^{(0)}_{n}}\right)
|\tilde\rho^{(0)}_{i}\rangle, \label{Rho1}
\end{eqnarray}
which leads to
\begin{eqnarray}
|\delta \tilde\rho_{n}\rangle &\equiv& |
\tilde\rho^{\exp}_{n}\rangle -|\tilde\rho^{(0)}_{n}\rangle \approx
|\delta \tilde\rho^{(1)}_{n}\rangle = \epsilon
|\tilde\rho^{(1)}_{n}\rangle. \label{delta_rho1}
\end{eqnarray}
Analogously, by using Eq.~(\ref{aux2}), one arrives at
\begin{eqnarray}
\langle\tilde\sigma^{(1)}_{n}| = -\sum_{i\;(i\neq n)}
\left(\frac{\langle\tilde\sigma^{(0)}_{n}| L_1
|\tilde\rho_{i}^{(0)}\rangle}{\lambda^{(0)}_{i}-\lambda^{(0)}_{n}}\right)
\langle\tilde\sigma^{(0)}_{i}|, \label{sigma1}
\end{eqnarray}
which leads to
\begin{eqnarray}
\langle\delta\tilde\sigma_{n}| \equiv
\langle\delta\tilde\sigma^{\exp}_{n}|
-\langle\delta\tilde\sigma^{(0)}_{n}| \approx
\langle\delta\tilde\sigma^{(1)}_{n} | \equiv \epsilon
\langle\tilde\sigma^{(1)}_{n}|, \label{delta_sigma1}
\end{eqnarray}
as derived in Ref.~\cite{James2001}.

\begin{figure}[htp!]
    \includegraphics[height=0.5\linewidth]{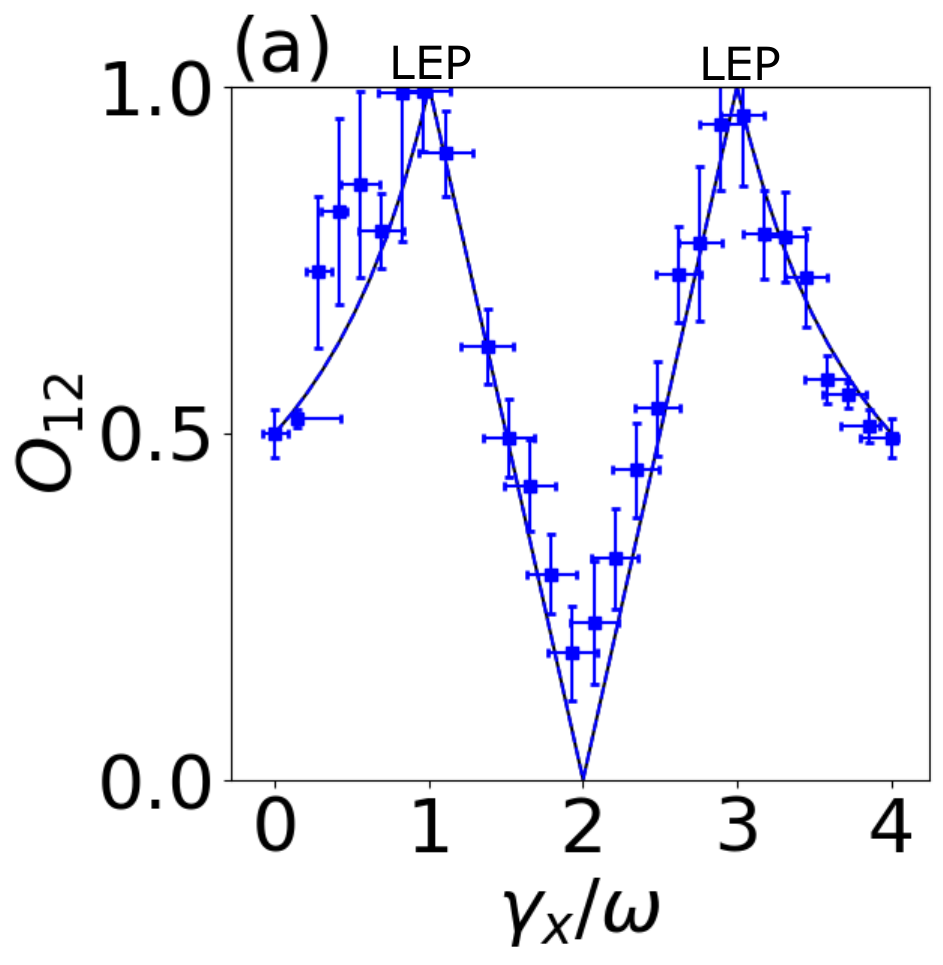}
    \includegraphics[height=0.5\linewidth]{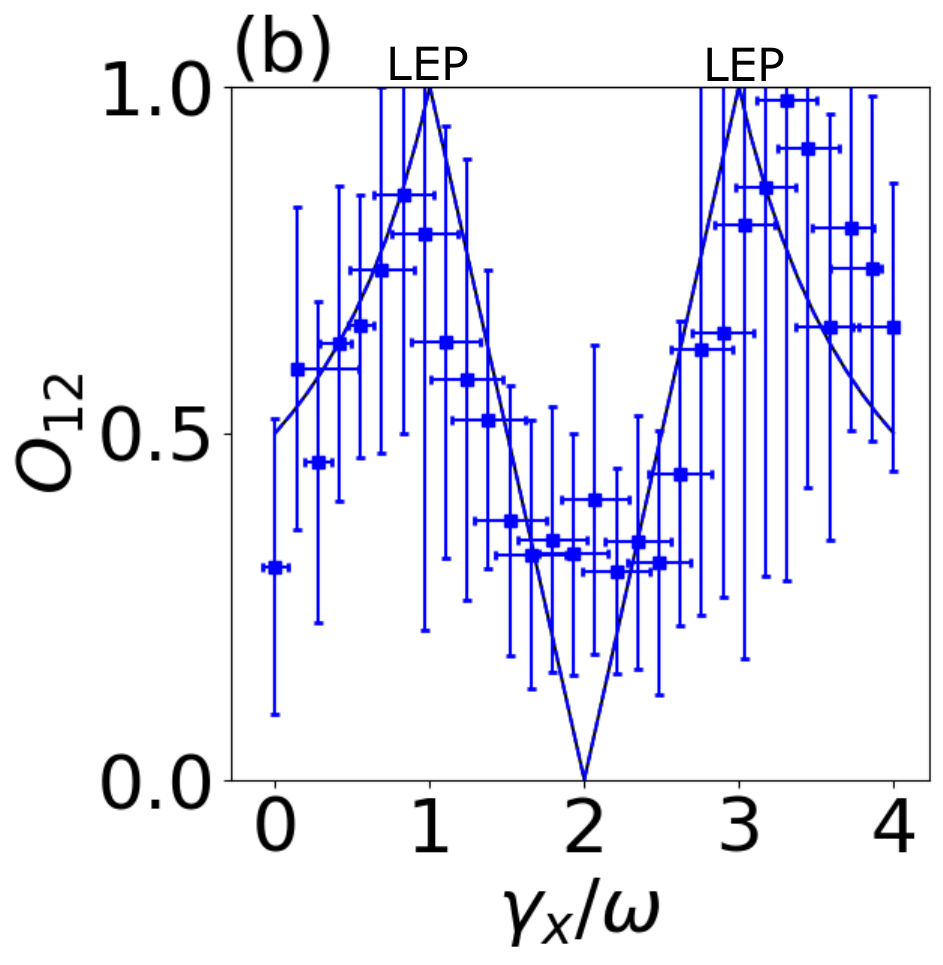}
\caption{Overlaps
$O_{12}=|\langle\tilde\sigma_{1}|\tilde\rho_{2}\rangle|$ of the
Liouvillian vectorized eigenmatrices $\langle \tilde\sigma_{1}|$
and $|\tilde\rho_{2}\rangle$ measured in our single- (a) and two-
(b) qubit experiments on an IBMQ processor (blue squares) and
compared to the corresponding theoretical predictions including
white noise (black curves). Same as Fig.~\ref{fig4}, but the error
bars $\delta \bar O^{\exp}_{12}$ are given here by
Eq.~(\ref{delta_S12bar}).}\label{fig8}
\end{figure}

Thus, the error bars $O^{\exp}_{12}$ of the scalar products
(overlaps)
$O_{12}^{\exp}=|\langle\tilde\sigma^{\exp}_{1}|\tilde\rho^{\exp}_{2}\rangle|$
of the experimental eigenmatrices $\langle
\tilde\sigma^{\exp}_{1}|$ and $|\tilde\rho^{\exp}_{1}\rangle$,
which are shown in Fig.~\ref{fig4}, are obtained as
\begin{eqnarray}
    \delta
O^{\exp}_{12}&=&\big|\langle\tilde\sigma^{\exp}_{1}|\tilde\rho^{\exp}_{2}\rangle- \langle\tilde\sigma^{(0)}_{1}|\tilde\rho^{(0)}_{2}\rangle\big| 
 \\
&=&\big|\langle\delta\tilde\sigma_{1}|\tilde\rho^{(0)}_{2}\rangle
 + \langle\tilde\sigma^{(0)}_{1}|\delta \tilde\rho_{2}\rangle+\langle\delta\tilde\sigma_{1}|\delta
 \tilde\rho_{2}\rangle\big|. \nonumber
 \label{delta_S12}
\end{eqnarray}
Note that these error bars are affected by the phase factors from
the overlaps, which can be arbitrary and depend on the applied
diagonalization method. Thus, to avoid this phase dependence, one
can redefine $\delta O^{\exp}_{12}$ as
\begin{equation}
    \delta \bar O^{\exp}_{12}= \Big[
 \big|\langle\delta\tilde\sigma_{1}|\tilde\rho^{(0)}_{2}\rangle\big|^2
 + \big|\langle\tilde\sigma^{(0)}_{1}|\delta \tilde\rho_{2}\rangle\big|^2
 + \big|\langle\delta\tilde\sigma_{1}|\delta \tilde\rho_{2}\rangle\big|^2
   \Big]^{1/2}.
 \label{delta_S12bar}
\end{equation}
These error bars are depicted in Fig.~\ref{fig8}, and can be
compared with those in Fig.~\ref{fig4}. Note that the point at
$\gamma_x=0$ in Fig.~\ref{fig4}(f) is fully consistent with our
theoretical prediction using this redefined error bar, $\delta
\bar S^{\exp}$, but it is not the case for $\delta S^{\exp}$.

\section{Estimation of errors in $\gamma/\omega$}

In our experiments, we intended to prepare quantum circuits to be
initially in pure states $|\psi(\gamma)\rangle,$ where the
dependence on $\gamma$ is given by the simulated quantum model.
However, due to random errors, the prepared states are not exactly
the intended pure states. Instead, we observe that the curves
associated with the prepared input states fit the experimental
data best if we assume the input state to be
\begin{equation}\label{eq:rho1}
\rho=(1-w)|\psi(\gamma)\rangle \langle \psi(\gamma)| +
\frac{w}{d}\openone_d,
\end{equation}
where $d$ is the dimension of the Hilbert space and $w$ stands for
the level of white noise. On the other hand, there are infinitely
many ways to decompose unity or to express the noisy input $\rho$
in a way that is compatible with our observations. In particular,
for a given $\gamma,$ the associated noisy state $\rho$ can be
expressed as
\begin{equation}\label{eq:rho2}
\rho=\int_{\gamma-\gamma_L}^{\gamma+\gamma_R} p(\gamma+\gamma')
|\psi(\gamma+\gamma')\rangle\langle \psi(\gamma+\gamma')|
d\gamma',
\end{equation}
where $p(\gamma)$ is a semipositive function defined on
$[\gamma-\gamma_L,\gamma+\gamma_R]$, such that $\rho$ is
normalized. To simplify the notation, let us set $\gamma = 0$ and
$\omega=1$ in our model. Both expressions for  $\rho$ should
result in the same fidelity with respect to the target input state
$|\psi(\gamma)\rangle$, i.e.,
\begin{equation}
     \int_{-\gamma_L}^{\gamma_R} P(\gamma')|\langle \psi(\gamma)|\psi(\gamma')\rangle
     |^2d\gamma'=1-w(1-1/d).
\end{equation}
Assume the integrand vanish for $\gamma\geq \gamma_R$ and
$\gamma\leq\gamma_L$:
\begin{equation}
      p(\gamma')|\langle \psi(\gamma)|\psi(\gamma')\rangle |^2=0 \quad \Rightarrow \hspace{0.5cm} p(\gamma_R)=p(-\gamma_L)=0.
\end{equation}
Let $p(\gamma)$ reach its maximum at $\gamma=0,$ which can be
interpreted as preparing the target state with a maximum
likelihood. If we assume that $p$ is a triangular distribution, we
arrive at
\begin{equation}
\hspace*{-2cm}
\int_{-\gamma_L}^0(a_L\gamma'+b)|\langle\psi(\gamma)|\psi(\gamma')\rangle|^2d\gamma'
+\int_0^{\gamma_R}(a_R\gamma'+b)|\psi(\gamma')\rangle|^2d\gamma'
=1-w(1-1/d),
\end{equation}
where $a_L=b/\gamma_L,$ $a_R=-b/\gamma_R,$ and $P(0)=b.$ We find
$b$ from the normalization condition and $\gamma_{L,R}$ are found
numerically from
\begin{eqnarray}
b\int_{-\gamma_L}^{\gamma_R}|\langle\psi(\gamma)|\psi(\gamma')\rangle|^2d\gamma'+\int_{-\gamma_L}^0a_L|\langle
\psi(\gamma)|\psi(\gamma')\rangle|^2\gamma'd\gamma'
\nonumber \\
+\int_0^{\gamma_R}
a_R|\langle\psi(\gamma)|\psi(\gamma')\rangle|^2\gamma'd\gamma'=1-w(1-1/d).\quad\quad
\end{eqnarray}
In this way, we estimated the maximum uncertainty in setting
$\gamma$. In general, we end up with $\gamma_L\neq \gamma_R,$
which corresponds to asymmetric uncertainties. We estimate the
left and right uncertainties (error bars for $\gamma$),
respectively, as for two independent triangular distributions to
be $P_L\gamma_L/\sqrt{6}$ and $P_R\gamma_R/\sqrt{6}$~\cite{NIST},
where
\begin{equation}
P_L = \int_{-\gamma_L}^0
p(\gamma')d\gamma'\qquad\mathrm{and}\qquad P_R =
\int_{0}^{\gamma_R} p(\gamma')d\gamma'.
\end{equation}

\section{Analytical formulas for the lossy driven qubit}

Here we show analytical results on the QPT of the lossy driven
qubit model analyzed in Sec.~III.

Case 1: Assuming   $\gamma_{-} = 0,$ we have
\begin{eqnarray}
 L'= \frac{\omega}{4}\left(
\begin{array}{cccccc}
 -4 & 4 & -1 & 1 & 0 & 0 \\
 4 & -4 & 1 & -1 & 0 & 0 \\
 1 & -1 & -2 x & 2 x & 0 & 0 \\
 -1 & 1 & 2 x & -2 x & 0 & 0 \\
 0 & 0 & 0 & 0 & -y_2 & y_2 \\
 0 & 0 & 0 & 0 & y_2 & -y_2 \\
\end{array}
\right), \label{L1}
\end{eqnarray}
where $x=\gamma/\omega$ and $y_k=2(x+k)$. The eigenvalues of $L'$
are:
\begin{eqnarray}
  \lambda'_1 &=& -2 \omega (2 + x), \nonumber \\
  \lambda'_2 &=& -\omega (2 + x - z), \nonumber \\
  \lambda'_3 &=& -\omega (2 + x + z),
\label{lambda1}
\end{eqnarray}
where $z=\sqrt{x^2-4 x+3}$, while the other three eigenvalues are
zero. The corresponding eigenmatrices are:
\begin{eqnarray}
  \ket{\tilde\rho'_1}&=&[0, 0, 0, 0, -1, 1]^{T}, \nonumber\\
  \ket{\tilde\rho'_2}&=&[2 - x - z, -2 + x + z, -1, 1, 0, 0]^{T},\nonumber\\
  \ket{\tilde\rho'_3}&=&[2 - x + z, -2 + x - z, -1, 1, 0, 0]^{T}.
\label{rho1}
\end{eqnarray}
Analogously, we find
\begin{equation}
 L'' = \omega \left(
\begin{array}{cccc}
 0 & 0 & 0 & 0 \\
 0 & -4 & -1 & 0 \\
 0 & 1 & -2 x & 0 \\
 0 & 0 & 0 & -y_2 \\
\end{array}
\right),
  \label{L2}
\end{equation}
for which the \emph{nonzero} eigenvalues of $L''$ are the same as
those of $L'$: $\lambda'_k=\lambda''_k$ ($k=1,2,3$), but their
eigenmatrices:
\begin{eqnarray}
  \ket{\tilde\rho''_1}&=&[0, 0, 0, 1]^{T},\nonumber\\
  \ket{\tilde\rho''_2}&=&[0, -2 + x + z, 1, 0]^{T},\nonumber\\
  \ket{\tilde\rho''_3}&=&[0, -2 + x - z, 1, 0]^{T},
\label{rho2}
\end{eqnarray}
are, in general, different from those of the corresponding
$\rho_k'$.

Case 2: Assuming $\gamma_{-} = \omega$, we have
\begin{equation}
 L'= \frac{\omega}{4} \left(
\begin{array}{cccccc}
 -9 & 9 & -2 & 2 & 2 & -2 \\
 9 & -9 & 2 & -2 & 2 & -2 \\
 2 & -2 & -4 x-1 & 4 x+1 & 2 & -2 \\
 -2 & 2 & 4 x+1 & -4 x-1 & 2 & -2 \\
 0 & 0 & 0 & 0 & -2y_2 & 2y_2 \\
 0 & 0 & 0 & 0 & 2y_3 & -2y_3 \\
\end{array}
\right),
  \label{L1case2}
\end{equation}
having the nonzero eigenvalues equal to:
\begin{eqnarray}
  \lambda'_1 &=& - \omega (2x + 5), \nonumber \\
  \lambda'_2 &=& -\frac{\omega}{2} (2x+5+2z), \nonumber \\
  \lambda'_3 &=& -\frac{\omega}{2} (2x+5-2z),
\end{eqnarray}
and the corresponding eigenmatrices:
\begin{eqnarray}
  \ket{\tilde\rho'_1}&=&\left[\bar x,\bar x,\bar x,\bar x, -\left(\frac{x+3}{x+2}\right), 1\right]^{T}, \nonumber\\
  \ket{\tilde\rho'_2}&=&[2-x-z, -2 + x + z, -1, 1, 0, 0]^{T},\nonumber\\
  \ket{\tilde\rho'_3}&=&[2 - x + z, -2 + x - z, -1, 1, 0, 0]^{T},
\label{veccase2}
\end{eqnarray}
where $\bar x=-1/[2(x+2)]$. Analogously, we find
\begin{equation}
 L'' = \frac{\omega}{2} \left(
\begin{array}{cccc}
 0 & 0 & -2 & 0 \\
 0 & -4x-1 & 0 & 2 \\
 0 & 0 & -4 x-10 & 0 \\
 0 & -2 & 0 & -9\\
\end{array}
\right).
  \label{LPcase2}
\end{equation}
As in Case~1, the \emph{nonzero} eigenvalues of $L''$ are the same
as those of $L'$: $\lambda'_k=\lambda''_k$ ($k=1,2,3$), but their
eigenmatrices are different as given by:
\begin{eqnarray}
  \ket{\tilde\rho''_1}&=&\left[\frac{1}{2x+5}, 0, 1, 1\right]^{T},\nonumber\\
  \ket{\tilde\rho''_2}&=&[0, -2 + x + z, 0, 1]^{T},\nonumber\\
  \ket{\tilde\rho''_3}&=&[0, -2 + x - z, 0, 1]^{T}.
\label{vecCase2}
\end{eqnarray}

\section{Measurement times}

Here we provide estimates for the total measurement times for the
different circuits.

1. For the experiments using the single-qubit circuit: The number
of experiments per data point was 72, the number of shots per
experiment was 20,000, and the time per shot was $5.7 \times
10^{-6}$~s. Thus, total number of shots was $N_{\rm total}=72 \,
{\rm experiments} \times 20,000 \, {\rm shots} = 1,440,000 \, {\rm
shots}$ and the total measurement time was $t_{\rm total} =
1,440,000 \, {\rm shots} \times 5.7 \times 10^{-6} \, {\rm s}
\approx 8.208 \, {\rm s}$.

2. For the experiments using the two-qubit circuit [shown in
Fig.~\ref{fig1}(b)]: The number of experiments per data point was
36, the number of shots per experiment was 20,000, and the time
per shot was $6.5 \times 10^{-6}$s. Thus, the total number of
shots was  $N_{\rm total}= 36 \, {\rm experiments} \times 20,000
\, {\rm shots} = 720,000 \, {\rm shots}$ and  the total
measurement time was $t_{\rm total} =  720,000 \, {\rm shots}
\times 6.5 \times 10^{-6} \, {\rm s} \approx 4.68 \, {\rm s}$

3. For the experiments using the three-qubit circuit [shown in
Fig.~\ref{fig1}(a)]: the number of experiments per data point was
18 and we performed 20,000 shots per experiment with the time per
shot about $8.9 \times 10^{-6}$s. Thus, the total number of shots
is given by $N_{\rm total}=18 \, {\rm experiments} \times 20,000
\, {\rm shots} = 360,000 \, {\rm shots}$, and the total
measurement time was $t_{\rm total} = 360,000 \, {\rm shots}
\times 8.9 \times 10^{-6} \, {\rm s} \approx 3.204 \,{\rm s}$

In summary, the estimated total measurement times are
approximately: 8.21s, 4.68s, and 3.20s for our experiments using
the single-, two-, and three-qubit circuits, respectively.

The same information can be expressed in the required numbers of
the quantum processor cycles. Here, we denote a single quantum
processor cycle as $\tau,$ instead of standard $dt,$ to
distinguish this time from the time scale of the studied dynamics.
In our experiments $\tau=0.2222\,{\rm ns}$. A single readout takes
$2552$ cycles, which accumulates to $0.11\,{\rm s}$ per
experiment. Including the measurement time, the execution times of
our quantum circuits are: $26\times 10^{3}\tau,$ $29\times
10^{3}\tau,$ and $40\times 10^{3}\tau$ for the single-, two- and
three-qubit experiments, respectively.

Note that an additional approx. $1-2$ seconds per experiment was
required for readout mitigation. This time was allocated for
applying dynamical decoupling (similar to the spin-echo method
commonly used in NMR) and calibrating signals to account for
detector imperfections. This added approximately 72 seconds for
the single-qubit experiments, 36 seconds for the two-qubit
experiments, and 18 seconds for the three-qubit experiments.

\vspace*{5mm}

\setlength{\parindent}{0pt}

{\bf DATA AVAILABILITY}

The experimental data and the Qiskit codes used to process the
data are available from the corresponding author upon reasonable
request.


\vspace*{2mm}

\noindent {\bf ACKNOWLEDGEMENTS}

The authors thank Anna Kowa\-lewska-Kud\l{}aszyk, Grzegorz
Chimczak, and Jan Pe\v{r}ina for insightful and useful
discussions. Our experiments were conducted on the IBM Quantum
Network through the IBM Quantum Hub operated by the Pozna\'n
Supercomputing and Networking Center (PSNC). The views expressed
are those of the authors, and do not reflect the official policy
or position of IBM or the IBM Quantum team. This work was
supported by the Polish National Science Centre (NCN) under the
Maestro Grant No. DEC-2019/34/A/ST2/00081. \c{S}.K.\"O.
acknowledges support from Air Force Office of Scientific Research
(AFOSR) Multidisciplinary University Research Initiative (MURI)
Award on Programmable systems with non-Hermitian quantum dynamics
(Award No. FA9550-21-1-0202).

\vspace*{2mm}

{\bf COMPETING INTERESTS}

The authors declare no competing interests.

\vspace*{2mm}

{\bf ADDITIONAL INFORMATION}

{\bf Supplemental document.} See Supplement~1, which includes
Ref.~\cite{AxlerBook}, for the derivations of the transformation
matrices for the three equivalent QPT methods.

\end{document}